\documentstyle[12pt]{article}

\newcommand{\eps}{\epsilon}

\newcommand{\beq}{\begin{equation}}
\newcommand{\eeq}{\end{equation}}
\newcommand{\lab}{\protect\label}
\newcommand{\cutoff}{\Lambda_\chi}

\begin{document}

\begin{titlepage}
\begin{flushright}
ROM2F 93/37\\
\end{flushright}
\vspace{0.5cm}
\begin{center}

{\Large\bf
Quark - Resonance Model}

\vspace*{1.0cm}

\vspace*{1.2cm}
         {\bf E. Pallante\footnotemark
      \footnotetext{ email: pallante@vaxtov.roma2.infn.it}} \\
{\footnotesize{ I.N.F.N., Laboratori Nazionali di Frascati,
Via E. Fermi, 00144 Frascati ITALY}} \\
\vspace*{0.5cm}
       and {\bf R. Petronzio \footnotemark \footnotetext{email:
petronzio@vaxtov.roma2.infn.it}}\\ {\footnotesize{Dipartimento di
Fisica, Universit\`a di Roma ``Tor Vergata'', Via della Ricerca
Scientifica,\\ 00133 Roma ITALY \\ and I.N.F.N., Sezione di Roma "Tor
Vergata".}}

\end{center}

\vspace*{0.5cm}

\begin{abstract}

We construct an effective Lagrangian for low energy hadronic
interactions through an infinite expansion in inverse powers of the low
energy cutoff $\Lambda_\chi$ of all possible chiral invariant
non-renormalizable interactions between quarks and mesons degrees of
freedom. We restrict our analysis to the leading terms in the $1/N_c$
expansion. The effective expansion is in $(\mu^2/\cutoff^2 )^P \ln
(\cutoff^2/\mu^2 )^Q$.  Concerning the next-to-leading order, we show
that, while the pure $\mu^2/\cutoff^2 $ corrections cannot be traced
back to a finite number of non renormalizable interactions, those of
order $(\mu^2/\cutoff^2 )
\ln (\cutoff^2/\mu^2 )$ receive contributions from a finite set of
$1/\cutoff^2$ terms. Their presence modifies the behaviour of observable
quantities in the intermediate $Q^2$ region. We explicitely discuss
their relevance for the two point vector currents Green's function.

\end{abstract}
\begin{flushleft}
ROM2F 93/37\\
December 93
\end{flushleft}
\end{titlepage}
\newpage

\section{Introduction.}

Effective chiral Lagrangians have become a relatively powerful technique
to describe hadronic interactions at low energy, i.e. below the chiral
symmetry breaking scale $\cutoff\simeq 4\pi f_\pi\sim 1$ GeV.  Chiral
perturbation theory (ChPt) \cite{GL1,GL2} describes the low energy
interactions among the pseudoscalar mesons $\pi , K, \eta $, which are
the lightest asymptotic states of the hadron spectrum and are identified
with the Goldstone bosons of the broken chiral symmetry.  The inclusion
of resonance degrees of freedom in the model (vectors, axials, scalars,
pseudoscalars and flavour singlets scalar and pseudoscalar) allows to
describe the interactions of all the particles below $\cutoff$
\cite{Fuji,Bando,Tens,PP}.  This approach has a disadvantage connected
with the non renormalizability of the effective low energy theory.
The chiral expansion (i.e. the expansion in powers of derivatives of
the low energy fundamental fields) results as an infinite sum over
chiral invariant operators of increasing dimensionality.  At each
order in the chiral expansion the number of terms increases.  The
coefficient of each term is not fixed by chiral symmetry and the
theory looses predictivity at higher orders.  Many attempts have been
done to reformulate the model in a more predictive fashion, both in
the non anomalous \cite{Tens} and in the anomalous sector
\cite{PP} of the theory.\par

In particular, there have been attempts to derive the low energy
effective theory from the more fundamental theory which describes the
interactions of quarks and gluons.  The first attempt to connect the low
energy effective theory of pseudoscalar mesons and resonances with QCD
has been proposed in \cite{ENJL}, where an application to strong
interactions of the old and well known Nambu-Jona Lasinio (NJL) model
\cite{NJL,Andrianov} is made.  The QCD Lagrangian is modified at long
distances (i.e. below the cutoff $\cutoff$) by an effective 4-quarks
interaction Lagrangian which remains chirally invariant.

The resonance and pseudoscalar mesons fields are introduced in the model
through the bosonization of the fermions effective action.

The ENJL model proposed in \cite{ENJL} includes only interaction terms
which are leading in an expansion in inverse powers of the cutoff
$\cutoff$.  This is a resonable approximation when we are interested in
the behaviour of the effective theory for light mesons at a very low
energy. Higher order terms bring powers of the derivative expansion term
$\partial / \cutoff$ which are indeed suppressed.

\noindent This is not the case in the intermediate and high energy region, i.e.
throughout the resonance region, where non renormalizable power-like
corrections arising from higher order terms can be dominant.  The ENJL
is not the full answer in the intermediate $Q^2$ region, while it can be
satisfactorily used to derive the effective Lagrangian of the
pseudo-Goldstone bosons (pions) at $Q^2=0$, where the resonance degrees
of freedom (effective at high $Q^2$) have been frozen out.

The paper is organized as follows.  In section 2 we construct the
quark-resonance model, which is the full effective model in the
intermediate $Q^2$ region.  In section 3 we derive explicitely all
possible higher dimensional terms of order $1/\cutoff^2$.  We explain
how these new vertices enter in the derivation of the effective meson
Lagrangian in section 4 by discussing in detail the case of the vector
resonance Lagrangian.  The presence of next-to-leading power -
leading-log corrections (NPLL), i.e. of order $Q^2/\cutoff^2 \ln
\cutoff^2/Q^2$, is crucial in the running of the effective Lagrangian
parameters at $Q^2\neq 0$.  The coefficients of the NPLL corrections
arising in the QR model should be fixed by experimental data.  In
section 5 we concentrate on the case of the two-point vector correlation
function, where we are able to extract significative informations on the
$Q^2$ behaviour of the real part of the invariant functions from the
existing data on the total $e^+e^-$ hadron cross section in the $I=1$
channel.  The results can be directly compared with the predictions
obtained in the ENJL framework \cite{2point,bijnens}. The corrections
improve the agreement with the experimental data significantly.

\section{ The model.}

The effective quark models describing low energy strong interactions
assume that the result of integrating over high frequency modes in the
original QCD Lagrangian, defined above a given energy cutoff, can be
expressed by additional non-renormalizable interactions.

For strong interactions the natural cutoff is the scale at which chiral
symmetry spontaneously breaks: $\Lambda_\chi\simeq 1$ GeV. The cutoff
sets the limit below which only the "low frequency modes" of the theory
are excited.

\noindent The QCD Lagrangian for the low frequency modes is modified as
follows:

\beq
{\cal L}_{QCD}\to {\cal L}_{QCD}^{\cutoff} + {\cal L}_{N.R.} (n-fermion).
\lab{CUT}
\eeq

\noindent ${\cal L}_{QCD}^{\cutoff}$ is the standard QCD Lagrangian where
only the low-frequency modes of quarks and gluons are present:

\beq
{\cal L}_{QCD}^{\cutoff} = \bar{q}(i\hat{D}-m_0)q.
\eeq

\noindent The second term in (\ref{CUT}) is the most general
non-renormalizable set of higher dimensional local n-fermion
interactions which respect the symmetries of the original theory and are
suppressed at low energy by powers of $Q^2/\cutoff^2$.

Recently, the Nambu- Jona Lasinio (NJL) model has been reanalysed in a
systematic way in the framework of hadronic low energy interactions
\cite{ENJL}.  Many applications and reformulations can be found in
\cite{MANYNJL}.

\noindent The extended version of the NJL model (ENJL) includes in
${\cal L}_{N.R.}(n-fermion)$ all the lowest dimension operators:
4-fermion local interactions which are leading in the $1/N_c$ expansion
(colour singlets) and respect all the symmetries of the original theory
(chiral symmetry, Lorentz invariance, P and C invariance). The form of
the effective Lagrangian is then uniquely determined:

\beq
{\cal L}^{ENJL} = {\cal L}_{QCD}^{\cutoff} + {\cal L}_{NJL}^{S,P} +
{\cal L}_{NJL}^{V,A} ,
\lab{4FERM}
\eeq

with

\beq
{\cal L}_{NJL}^{S,P} = {8\pi^2 G_S(\cutoff )\over N_c\cutoff^2}\sum_{a,b}
(\bar{q}_R^aq_L^b)(\bar{q}_L^bq_R^a)
\lab{4FERMS}
\eeq

and

\beq
{\cal L}_{NJL}^{V,A} = {8\pi^2 G_V(\cutoff )\over
N_c\cutoff^2}\sum_{a,b}
\bigl [ (\bar{q}_L^a\gamma_\mu q_L^b)(\bar{q}_L^b\gamma^\mu q_L^a)
+ L\to R\bigr ] .
\lab{4FERMV}
\eeq

The {\em current quarks} $q_{L,R}$ transform as $q_{L,R}\to
g_{L,R}q_{L,R}$ under the chiral flavour group $SU(3)_L\times SU(3)_R$,
with elements $g_{L,R}$.  As pointed out in \cite{ENJL} the 4-quark
effective vertex can be thought as a remnant of a single "low frequency"
gluon exchange diagram (see fig. 1).  The gluon propagator modified at
high energy with a cutoff

\beq
{1\over Q^2}\to \int^{1\over\cutoff^2}_0~d\tau e^{-\tau Q^2}
\eeq

\noindent leads to a local effective 4-quark interaction

\beq
{g_s^2\over\cutoff^2} (\bar{q}\gamma_\mu{\lambda^{(a)}\over 2}q)
(\bar{q}\gamma_\mu{\lambda_{(a)}\over 2}q).
\eeq

\noindent
By means of the Fierz-identities one gets the $S,P,V,A$ combinations of
(\ref{4FERMS},\ref{4FERMV}) with the identification $G_S=4G_V$.

The non-renormalizable part of the fermion action $S_{NR}(q)$ admits
an integral representation in terms of {\em auxiliary boson fields}:

\beq
e^{iS_{NR}[q]} = \int {\cal D}B~e^{iS[B,q]}.
\eeq

The previous relation introduces the meson degrees of freedom into the
effective quark Lagrangian. The following two identities hold:

\begin{eqnarray}
\exp{i\int d^4x~{\cal L}_{S,P}(x)}&=& \int {\cal D}
H~\exp {i\int d^4x\biggl\{-(\bar{q}_LH^\dagger
q_R+h.c.)-{N_c\cutoff^2\over 8\pi^2G_S}~tr (HH^\dagger)\biggr\}}
\nonumber\\
\exp{i\int d^4x~{\cal L}_{V,A}(x)}&=& \int {\cal D}L_\mu {\cal D}
R_\mu~\exp i\int d^4x\biggl\{ \bar{q}_L\gamma_\mu L^\mu q_L \nonumber\\
&&+{N_c\cutoff^2\over 8\pi^2G_V}~{1\over 4}~tr (L_\mu L^\mu )+
L\to R \biggr\} ,
\end{eqnarray}

\noindent
where we have introduced three auxiliary fields: a scalar field $H(x)$
and the right-handed and left-handed fields $L_\mu$ and $R_\mu$.  Under
the chiral group they transform as:

\begin{eqnarray}
H&\to& g_R H g_L^\dagger \nonumber\\
L_\mu&\to& g_L L_\mu g_L^\dagger \nonumber\\
R_\mu&\to& g_R R_\mu g_R^\dagger .
\end{eqnarray}

\noindent
The field $H$ can be decomposed into the product of a new scalar field
$M$ times a unitary field $U$:

\beq
H = MU = \xi\tilde{H}\xi ,
\eeq

\noindent
where the field $\xi$ is the square root of the field $U$: $\xi^2 = U$.
A connection with the physical fields is obtained by redefining the
auxiliary fields as follows:

\begin{eqnarray}
H&=&\xi\tilde{H}\xi\nonumber\\
W^+_\mu &= &\xi L_\mu\xi^\dagger +\xi^\dagger R_\mu \xi \nonumber\\
W^-_\mu &=& \xi L_\mu\xi^\dagger -\xi^\dagger R_\mu \xi
\end{eqnarray}

\noindent
The new set of fields transforms homogeneously under chiral transformation:

\beq
\bigl\{\tilde{H},W^+_\mu ,W^-_\mu\bigr\}\to h\bigl\{\tilde{H},W^+_\mu ,
W^-_\mu\bigr\}h^\dagger ,
\eeq

\noindent where $h$ is a non linear representation of the chiral group.

\noindent
We redefine also the fermion fields by replacing the {\em current quarks}
$q_{L,R}$ with the  {\em constituent quarks}:

\beq
Q_L = \xi q_L~~~~~~~~~~~~~Q_R = \xi^\dagger q_R .
\eeq

\noindent They transform under the chiral group $G = SU(3)_L\times SU(3)_R$
as:

\beq
Q_L\to h(\Phi, g_L,g_R )Q_L~~~~~~~~~~~Q_R\to h(\Phi,g_L,g_R )Q_R,
\eeq

\noindent
where the matrix $h(\Phi ,g_L,g_R )$ acts on the element $\xi$ of the coset
group $G/SU(3)_V$

\beq
\xi (\Phi ) \to g_R \xi (\Phi )h^\dagger  =h\xi (\Phi ) g_L^\dagger .
\eeq

\noindent The quark field Q is defined as $Q=Q_L+Q_R$.

In terms of the new variables the bosonized euclidean generating functional
of the ENJL model reads:

\begin{eqnarray}
&&Z[v,a,s,p]= \int ~{\cal D}\xi ~{\cal D}{\tilde H}~ {\cal D}L_\mu
{}~{\cal D} R_\mu ~e^{-\Gamma_{eff}[\xi ,W^+_\mu ,W^-_\mu ,{\tilde H} ;
{}~v,a,s,p]}
\nonumber\\
&&\nonumber\\ &&e^{-\Gamma_{eff}[\xi ,W^+_\mu ,W^-_\mu ,{\tilde H} ;
{}~v,a,s,p]}= \nonumber\\ &&\exp~ \biggl (-\int d^4 x \biggl\{
{N_c\cutoff^2\over 8\pi^2 G_S(\cutoff )} tr {\tilde H}^2 +
{N_c\cutoff^2\over 8\pi^2 G_V(\cutoff )}{1\over 4} tr (W^+_\mu W^{+\mu}
+ W^-_\mu W^{-\mu} )\biggr\}\biggr )\times \nonumber\\ &&{1\over Z}\int~
{\cal D}G_\mu \exp~ \biggl (-\int d^4x {1\over 4}
G_{\mu\nu}^{(a)}G^{(a){\mu\nu}} \biggr )
\int~ {\cal D}Q {\cal D}\bar{Q}\exp~ \biggl (\int d^4 x\bar{Q}D_E Q \biggr ),
\lab{ENJLBOS}
\end{eqnarray}

\noindent where we have defined the total differential operator $D_E$ as
follows:

\beq
D_E = \gamma_\mu\nabla_\mu -{1\over 2}(\Sigma-\gamma_5\Delta )- {\tilde
H}(x),
\eeq

\noindent with the covariant derivative acting on the chiral quark field
given by:

\beq
\nabla_\mu = \partial_\mu + iG_\mu +  \Gamma_\mu -{i\over 2} W^+_\mu
-{i\over 2} \gamma_5 (\xi_\mu - W^-_\mu ).
\eeq

\noindent The field $\Gamma_\mu$ acts like a vector field and is defined by:

\beq
\Gamma_\mu= {1\over 2}\{\xi^\dagger d_\mu\xi + \xi d_\mu\xi^\dagger \}
={1\over 2} \{ \xi^\dagger [\partial_\mu - i(v_\mu + a_\mu )]\xi + \xi
[\partial_\mu - i(v_\mu - a_\mu )] \xi^\dagger \}.
\lab{GMU}
\eeq

\noindent
It transforms inhomogeneously under the local vector part of the chiral group

\beq
\Gamma_\mu\to h\Gamma_\mu  h^\dagger  +
h \partial_\mu  h^\dagger
\eeq

\noindent
and makes the derivative on the $Q$ field invariant under local vector
transformations.

\noindent The field $\xi_\mu$ is like an axial current and is defined by:

\begin{eqnarray}
\xi_\mu&=&i\{ \xi^\dagger d_\mu\xi - \xi d_\mu\xi^\dagger \}\nonumber\\
&=& i\{ \xi^\dagger [\partial_\mu - i(v_\mu + a_\mu )]\xi - \xi
[\partial_\mu - i(v_\mu - a_\mu )] \xi^\dagger \} = \xi_\mu^\dagger .
\end{eqnarray}

\noindent It transforms homogeneously under the chiral group $G$:

\beq
\xi_\mu\to h\xi_\mu  h^\dagger .
\eeq

\noindent The fields $\Sigma$ and $\Delta$ are defined by:

\begin{eqnarray}
\Sigma&=&\xi^\dagger {\cal M} \xi^\dagger +\xi {\cal M}\xi \nonumber\\
\Delta&=&\xi^\dagger {\cal M} \xi^\dagger -\xi {\cal M}\xi .
\end{eqnarray}

\noindent They are both proportional to the quark mass matrix ${\cal M}$ and
vanish in the chiral limit.  The field ${\tilde H}(x)$ is the auxiliary
scalar field of the bosonized action and can be parametrised as

\beq
{\tilde H}(x) = M_Q{\bf 1} +\sigma (x),
\eeq

\noindent where we have split the ${\tilde H}$ field into its vacuum
expectation value and the fluctuation around it. The quantity $M_Q$ is
the value of the ${\tilde H}(x)$ field (used in the so called mean field
approximation of the ENJL model) which minimizes the effective action in
absence of other external fields:

\beq
{\delta\Gamma_{eff}({\tilde H},..)\over \delta{\tilde H}}\vert_{
\xi = 1 ,W^+_\mu =W^-_\mu =0 ;v,a,s,p =0; {\tilde H}=<{\tilde H}>} = 0.
\lab{MGAP}
\eeq

\noindent $M_Q\neq 0$ corresponds to broken chiral symmetry \cite{boh}.
Its value is the solution of the mass gap equation generated by
(\ref{MGAP}).

In the leading effective action (\ref{ENJLBOS}) two constants appear:
the scalar coupling $G_S$ and the vector coupling $G_V$. They are
functions of the cutoff $\cutoff$ and their estimate involves
non-perturbative contributions.\par

The fundamental fields of the bosonized action are
$\xi_\mu ,\Gamma_\mu , W^+_\mu ,W^-_\mu , \tilde{H}$.

\par\noindent A full effective quark model {\em \`a la NJL} contains a
priori an infinite tower of n-fermion operators with increasing
dimensionality: the ENJL 4-fermion interactions are the leading terms
both in $1/\cutoff$ and $1/N_c$ expansions.

The QR model is the bosonization of the full effective quark model
{\em \`a la NJL}.

\par\noindent The resulting quark-resonance Lagrangian is a non-renormalizable
Lagrangian which contains all possible interaction terms between quarks
and resonances.  The QCD euclidean generating functional of the
correlation functions at low energy is modified as follows:

\begin{eqnarray}
Z[v,a,s,p]&=&e^{W[v,a,s,p]} = \int~ {\cal D}\Gamma ~e^{-\Gamma_{eff}
[\Gamma ;v,a,s,p]} ~e^{-f[\Gamma ]},
\lab{27}
\end{eqnarray}

\noindent where $\Gamma$ indicates the set of fields introduced in the
bosonization of the QCD effective action and the effective low energy
action $\Gamma_{eff}$ is given by

\begin{eqnarray}
e^{-\Gamma_{eff} [\Gamma ;v,a,s,p]} &=& {1\over Z}\int {\cal D}G_\mu \exp~
\biggl (-\int d^4x
{1\over 4}G_{\mu\nu}^{(a)}G^{(a){\mu\nu}} \biggr )
\int {\cal D}Q {\cal D}\bar{Q}\nonumber\\
&&~\exp~ \biggl [ \int d^4 x~
\biggl ( \bar{Q}\gamma^\mu (\partial_\mu +i G_\mu )Q +\sum_0^\infty
\biggl ({1\over\cutoff}\biggr )^n
\bar{Q}\Gamma Q \biggr )\biggr ].
\lab{LAG}
\end{eqnarray}

\noindent The functional $f[\Gamma ]$ in eq. (\ref{27})
contains the mass terms of the auxiliary boson fields.

The second term in (\ref{LAG}) is the non-renormalizable part of the
action expressed as an infinite sum over all chiral invariant quark
bilinears interacting with the low energy boson degrees of freedom.  The
most general structure of the $\Gamma$ operator can be represented by:

\beq
\Gamma = \beta (\cutoff )\times \{\gamma_{Dirac}\}
\times \{ \Gamma_\mu , \xi_\mu , W^+_\mu , W^-_\mu , \sigma \}
\times d_\mu^{~n} ,
\eeq

where the coupling $\beta (\cutoff )$ is not deducible from symmetry
principles and not calculable in a perturbative way.  $d_\mu$ defines
the covariant derivative acting on the meson fields or on the chiral
quark fields and the set $\{\Gamma_\mu , \xi_\mu , W^+_\mu , W^-_\mu ,
\sigma \}$ contains all possible boson fields which can couple to the
quark bilinears and which can be identified with the physical degrees of
freedom of the low energy effective theory: pseudoscalar mesons and
resonances.  As it is shown in detail in ref. \cite{ENJL}, the
integration over quark fields induces a mixing between the axial field
$W^-_\mu$ and the pseudoscalar field $\xi_\mu$ and a diagonalization is
required to define the true physical axial and pseudoscalar meson
fields.  Our Lagrangian at leading order in the $1/\cutoff$ expansion
and in the $1/N_c$ expansion coincides with the bosonization of the ENJL
model of eq. (\ref{ENJLBOS}).

The additional quark-meson interaction terms originate from the
bosonization of non-renormalizable n-quark vertices. These can be of two
types:

\begin{eqnarray}
(I)&& \hbox{4-quark}\times \hbox{derivatives} \nonumber\\ (II)&& \hbox{n
($>$4) - quark}\times \hbox{derivatives}
\end{eqnarray}

\noindent The higher dimension terms
with n-quarks can be easily constructed using the chiral invariant
building blocks of the ENJL model:

\begin{eqnarray}
&&\bar{q}\hat{D}q=\bar{q}_L\hat{D}q_L+\bar{q}_R\hat{D}q_R \nonumber\\
&&(\bar{q}q)^2-(\bar{q}\gamma_5 q)^2 = 4\bar{q}_Lq_R\bar{q}_Rq_L =
S^2-P^2 \nonumber\\ &&(\bar{q}\gamma_\mu q)^2+(\bar{q}\gamma_\mu\gamma_5
q)^2 = 2[(\bar{q}_L\gamma_\mu q_L)^2 +(\bar{q}_R\gamma_\mu q_R)^2] = V^2
+ A^2
\lab{BB}
\end{eqnarray}

\noindent
There is one chiral invariant term with four quarks, which does not appear in
the ENJL model:

\beq
{\cal L}_\sigma = C (\bar{q}^a\sigma_{\mu\nu}q^b)(\bar{q}_b\sigma^{\mu\nu}q_a).
\eeq

\noindent This term is a magnetic moment like term and contributes to
the $g-2$ of the muon, which however is next-to-leading (i.e.
${\cal{O}}(1)$) in the $1/N_c$ expansion \cite{Peris} and will be
neglected.

 \noindent All possible higher order terms can be constructed as
combinations of arbitrary powers of the basic building blocks of eq.
(\ref{BB}) times the insertion of powers of the differential operator
$\partial^2/\cutoff^2$.

The bosonization of the most general n-fermion action ends up in the
most general quark-resonance action (\ref{LAG}) constrained only by
chiral symmetry.

\noindent The possible relevance of additional non-renormalizable terms
in the scalar sector of the NJL model has been already pointed out in
\cite{Andrianov}.  They modify the mass-gap equation and can be
incorporated in a renormalization of the scalar coupling $G_S$, or
alternatively of the expectation value of the scalar field $M_Q$ which
minimizes the effective potential.

The effective meson theory is given by the integral over quarks and
gluons of the Lagrangian (\ref{LAG}). By neglecting the gluon
corrections, which are inessential to our argument, the derivation of
the low energy theory reduces to the integral over quarks of the
quark-resonance effective Lagrangian:

\begin{eqnarray}
&&\int\int {\cal D}Q {\cal D}\bar{Q}~\exp~ \biggl [ \int d^4 x~
\biggl ( \bar{Q}\gamma^\mu (\partial_\mu +i G_\mu )Q +\sum_0^\infty
\biggl ({1\over\cutoff}\biggr )^n
\bar{Q}\Gamma Q \biggr )\biggr ]  \nonumber\\
&=& \det \biggl [ {\hat {D_0}} +
\sum_0^\infty \biggl ({1\over\cutoff}\biggr )^n\Gamma \biggr ] ,
\end{eqnarray}

\noindent where  ${\hat {D_0}} = \gamma^\mu (\partial_\mu +i G_\mu )$
is the free fermion operator.  The integral corresponds to the set of
all one quark-loop diagrams which mediate the interactions among the
meson fields as shown in fig. 2.  Higher order terms contain factors of
$\partial^2/\cutoff^2$. The derivative acts: a) on the quark in the
loop, b) on the external boson legs.  In both cases it produces powers
of the external momenta in the diagram.

In order to clarify how these contributions enter in the effective meson
Lagrangian, we consider the one quark-loop diagram constructed with the
insertion of two leading vertices involving the vector field
$\bar{Q}\gamma_\mu W^+_\mu Q$.  This diagram gives the leading
contribution to the vector wave function.

\noindent The next-to-leading diagram is given by the insertion of a
next-to-leading vertex of the form ${1\over\cutoff}\bar{Q}\gamma_\mu
W^+_\mu d^2 Q$ and a leading vertex.  This gives next-to-leading
corrections to the vector wave function which contain powers of the
ratio ${Q^2\over\cutoff^2}$.  The $Q^2$ dependence of next-to-leading
terms affects the running in energy of the parameters of the effective
meson Lagrangian.

In what follows we concentrate on the next-to-leading vertices (up to
$1\over\cutoff^2$) of the quark-resonance model and on their
contributions to the parameters of the vector resonance Lagrangian which
are leading in the chiral expansion.

\noindent The analysis shows two basic features:

\begin{itemize}
\item
Next-to-leading corrections increase the number of independent
parameters present at the leading order, so that higher order
contributions cannot be reabsorbed in a redefinition of the leading
parameters.  This implies that relations among resonance parameters
valid at zero energy (i.e. at the leading order) are modified when the
energy increases (i.e. including next-to-leading corrections).

\item
There are two types of next-to-leading corrections:

\end{itemize}

\indent\indent
i) NPLL corrections $={Q^2\over \cutoff^2}~\ln{\cutoff^2\over Q^2}$

\indent\indent
ii) NTL power corrections without logarithms (NP) $={Q^2\over \cutoff^2}~1$.

\vspace{0.8cm}
The first class receives contribution from a finite set of higher
dimension operators (only $1\over\cutoff^2$ terms) and can be easily
kept under control.

\noindent Chiral symmetry does not constrain the coefficients
$\beta (\cutoff )$ and one has to fix them from experimental data.
We defer to section (\ref{VV}) a discussion of the particular
case of the vector 2-point function.

\section{The Lagrangian up to $1\over \Lambda_\chi^2$ order.}

\indent The invariant quark-resonance bilinears up to
$1\over \Lambda_\chi^2$ order are all the possible quark-resonance
bilinears which are chirally invariant and respect all the usual
symmetries: Parity, Charge conjugation and Lorentz invariance.  We
report in Table 1 and Table 2 the P and C transformation properties of
the quark bilinears and the fundamental fields in the $\Gamma$ set
respectively.  If we are interested in producing the effective
Lagrangian of meson resonances at leading order in the chiral expansion
only two classes of quark-resonance bilinears give contribution:

\begin{eqnarray}
(I)&&  \biggl ({1\over\cutoff}\biggr )^n\times 1~b\times n~d \nonumber\\
(II)&&\biggl ({1\over\cutoff}\biggr )^n\times 2~b\times (n-1)~d ,
\end{eqnarray}

\noindent with $b=$boson field and $d=$covariant derivative.
These are also the only possible terms that appear at order
$1/\cutoff^2$.

\noindent We work in the chiral limit and
we set to zero all terms proportional to the mass fields $\Sigma$ and
$\Delta$.  As it is shown in detail in ref. \cite{ENJL} the integration
over quarks induces a mixing between the pseudoscalar field $\xi_\mu$
and the axial field $W^-_\mu$ which is leading in the chiral expansion.
The physical fields are obtained after a diagonalization of the
quadratic matrix. In the ENJL model this corresponds to a rescaling of
the pseudoscalar field by the mixing parameter $g_A$, which the authors
of ref. \cite{ENJL} connect to the $g_A$ parameter of the effective
quark-model by Georgi-Manohar \cite{Georgi}.  In the QR model the mixing
parameter $g_A$ receives higher order corrections: the physical
pseudoscalar field is given by the rescaling:

\beq
\xi_\mu\to g^\prime_A\xi_\mu
\eeq

\noindent with the new mixing parameter $g^\prime_A$. Each insertion of a
field $\xi_\mu$
brings an insertion of the mixing parameter $g^\prime_A$.

The leading terms of the ENJL model have a logarithmic dependence upon
the cutoff $\cutoff$. Terms without logarithms can receive contributions
from all higher order terms. Indeed, besides the finite contributions of
the leading renormalizable operators, higher dimensions
non-renormalizable operators differing from the leading ones by powers
of derivatives may develop divergences that, integrated up to the cutoff
$\cutoff$, compensate the inverse powers of $\cutoff$ and contribute as
constant terms. The same happens to the terms which are of order
$1/\cutoff^2$ in the final low energy meson Lagrangian: those
accompanied by logarithms can be traced back to terms of order
$1/\cutoff$ and $1/\cutoff^2$ in the original quark-resonance Lagrangian
while those without logarithms are determined by the whole tower of
non-renormalizable interactions.

At the order ${1\over\cutoff}$ we have the following terms:

\begin{eqnarray}
{1\over\cutoff}\times 1~b\times 1d &:& {1\over\cutoff}\times
\bar{Q}\biggl [\sigma_{\mu\nu}W^{+\mu\nu}
+ \sigma_{\mu\nu}(W^{+\mu} d^\nu - W^{+\nu} d^\mu )\biggr ] Q,
\nonumber\\
&&{1\over\cutoff}\times \bar{Q}\sigma_{\mu\nu}\Gamma^{\mu\nu}Q,
\nonumber\\ {1\over\cutoff}\times 2~b&:& {1\over\cutoff}\times
\bar{Q}\biggl [ W^+_\lambda W^{+\lambda} +
\sigma_{\mu\nu}[W^{+\mu} ,W^{+\nu} ]\biggr ] Q,  \nonumber\\
&&{1\over\cutoff}\times \bar{Q}\biggl [ \xi_\lambda \xi^\lambda  +
\sigma_{\mu\nu}[\xi^\mu ,\xi^\nu ] \biggr ] Q,
\end{eqnarray}

\noindent where
$\Gamma^{\mu\nu}$ is defined as $\Gamma_{\mu\nu}=\partial_\mu\Gamma_\nu
-\partial_\nu\Gamma_\mu +[\Gamma_\mu , \Gamma_\nu ]$.  Terms containing
the combination $(\Gamma^\mu d^\nu - \Gamma^\nu d^\mu )$ with
derivatives acting on quark fields are forbidden by chiral invariance
because of the inhomogeneous transformation of the $\Gamma_\mu$ field.
The $\sigma_{\mu\nu}$ terms come from the bosonization of a 4-fermion
vertex like the one in (\ref{4FERMV}) with $\gamma_\mu$ replaced by the
tensor $\sigma_{\mu\nu}$ and are negligible in a leading $N_c$
expansion.

All possible invariants at $1/\cutoff^2$ order are terms with one and
two meson fields. For semplicity, we write explicitely only the terms
containing the fields $W^+_\mu$ (vector resonance) and $\xi_\mu$ (the
axial current of pseudoscalar mesons), which we will use in the next
section. They are:

\begin{eqnarray}
{1\over\cutoff^2}\times 1~b\times 2d &:& {1\over\cutoff^2}\times
\bar{Q}\biggl [
\gamma_\mu d_\lambda W^{+\mu\lambda} + \gamma_\mu  W^{+\mu\lambda} d_\lambda
+ \gamma_\mu (W^{+\mu} d^\lambda - W^{+\lambda} d^\mu ) d_\lambda \biggr ] Q,
\nonumber\\
&&{1\over\cutoff^2}\times  \bar{Q} \biggl [
\gamma_\mu \gamma_5d_\lambda \xi^{\mu\lambda} + \gamma_\mu \gamma_5
\xi^{\mu\lambda} d_\lambda
+ \gamma_\mu\gamma_5 (\xi^\mu d^\lambda - \xi^\lambda d^\mu ) d_\lambda
\biggr ] Q,  \nonumber\\
{1\over\cutoff^2}\times 2~b\times 1d &:& {1\over\cutoff^2}\times
\bar{Q} \biggl [
\gamma_\mu [W^+_\nu , W^{+\mu\nu}] + \gamma_\mu [W^+_\nu , W^{+\mu} d^\nu
- W^{+\nu} d^\mu \biggr ] Q, \nonumber\\
&& {1\over\cutoff^2}\times \bar{Q} \biggl [
\gamma_\mu [\xi_\nu , \xi^{\mu\nu}] + \gamma_\mu [\xi_\nu , \xi^\mu d^\nu
- \xi^\nu d^\mu \biggr ] Q, \nonumber\\
&&{1\over\cutoff^2}\times  \bar{Q} \biggl [
\gamma_\mu\gamma_5 [\xi_\nu , W^{+\mu\nu}] +
\gamma_\mu\gamma_5 [W^+_\nu , \xi^{\mu\nu}] +
\gamma_\mu\gamma_5 [\xi_\nu , W^{+\mu} d^\nu - W^{+\nu} d^\mu ] \nonumber\\
&& +
\gamma_\mu\gamma_5 [W^+_\nu , \xi^\mu d^\nu - \xi^\nu d^\mu ] \biggr ] Q.
\end{eqnarray}

\noindent To
these terms we have to add all the analogous terms with the
substitutions $W^+_{\mu\nu}\to\Gamma_{\mu\nu}$ and $\xi_\mu\to A_\mu$,
containing the axial resonance field $A_\mu$ and the vector current of
the pseudoscalar mesons $\Gamma_\mu$.  In the next section we derive the
next to leading corrections to the parameters of the leading chiral
vector resonance Lagrangian.

\section{ The Vector Meson Lagrangian.}

\indent The leading
non anomalous Lagrangian with one vector meson ${\cal L}_V$
(i.e. of order $p^3$) is the following:

\begin{eqnarray}
{\cal L}_V &=& -{1\over 4}<V_{\mu\nu}V^{\mu\nu}> +{1\over 2} M_V^2
<V_\mu V^\mu > - {f_V\over 2\sqrt{2}} <V_{\mu\nu}f^{\mu\nu}_+>
\nonumber\\ &&-i{g_V\over 2\sqrt{2}} <V_{\mu\nu}[\xi^\mu ,\xi^\nu ]>
+H_V <V_\mu [\xi_\nu ,f^{\mu\nu}_- ]> +iI_V <V_\mu [\xi^\mu ,\chi_- ]>.
\nonumber\\&&
\lab{LAGV}
\end{eqnarray}

The form above corresponds to the so called Conventional Vector model
\cite{Tens,PP} where the vector fields are introduced as ordinary
fields. This is the natural form for the effective low energy theory
after the bosonization of four-fermion interactions.  In the chiral
limit the $I_V$ term is zero and the Lagrangian is parametrised by five
constants: the vector resonance wave function $Z_V$, the mass $M_V$ and
the coupling constants $f_V, g_V$ and $H_V$.\par The ENJL estimate of
the five parameters has been already derived in \cite{ENJL,PRADES} using
the heat kernel expansion technique in the calculation of the fermion
determinant. We rederive both the leading and non-leading contributions
using the loopwise expansion.  If we write the fermion differential
operator as a sum of a free part $D_0$ and of a perturbation $\delta$,
the euclidean effective action at one loop is given by:

\beq
\Gamma_{eff}(\delta )
=-Tr\ln~[D_0+\delta ]+ Tr\ln~D_0 = - TrD_0^{-1}\delta
+{1\over 2} Tr D_0^{-1}\delta D_0^{-1}\delta + ....,
\eeq

\noindent where we have subtracted its value at $\delta =0$.

The various terms on the r.h.s. are identified by the order n in the
series expansion of the logarithm.  The term $TrD_0^{-1}\delta$ (n=1)
contains the tadpole graphs. The next term (n=2) contains the set of
graphs with the insertion of two vertices in the loop.  The
contributions to the parameters of ${\cal L}_V$ arise from the $n=2$ and
$n=3$ insertions of vertices in the perturbative expansion,
corresponding to the insertion of three meson fields at most.

\noindent At the leading order and in the chiral limit
$\delta$ is given by:

\beq
\delta = \delta_0 =
\gamma_\mu [\Gamma_\mu -{i\over 2}W^+_\mu -{i\over 2}\gamma_5 (\xi_\mu-
A_\mu )],
\lab{DELTA0}
\eeq

\noindent and the free part $D_0$ is

\beq
D_0 = \gamma_\mu (\partial_\mu +iG_\mu -M_Q).
\eeq

\noindent The mass term $M_Q$ acts as an infrared cutoff in quark loop
diagrams.

\noindent The complete operator $\delta$ is the sum of the leading
part $\delta_0$ defined in (\ref{DELTA0}) and the non leading
contributions in the $1/
\cutoff$ expansion:

\beq
\delta = \delta_0 + \sum_{n=1}^\infty\biggl ({1\over \cutoff}\biggr )^n
\Gamma .
\lab{DELTATOT}
\eeq

In Appendix A the single quark-loop diagrams for n=2 are explicitely
calculated with the insertion of a generic form of the operator $\delta
(x)$.  Using the formulas reported there one can get the contribution to
a given parameter of the vector Lagrangian with the substitution of the
appropriate operator $\delta (x)$. The next order (n=3) is a
straightforward generalization of the previous case and it will not
enter in the calculation of the parameters that are analysed in detail
in sections 4.3 and 5.

We distinguish two classes of terms in the ${\cal L}_V$ Lagrangian: the
kinetic term $Z_V$, the mass term $M_V$ and the interaction term $f_V$
belong to the first class and contain two meson fields plus derivatives,
while the $g_V$ and $H_V$ terms belong to the second class and contain
three fields plus derivatives.  The $H_V$ term is a three fields term
because of the identity $f_{\mu\nu}^-=\xi_{\mu\nu}= d_\mu\xi_\nu -
d_\nu\xi_\mu$, while the $g_V$ term belongs also to the two fields class
because of the identity

\beq
\Gamma_{\mu\nu} = -{i\over 2}f_{\mu\nu}^+ +{1\over 4}[\xi_\mu ,\xi_\nu ].
\lab{GVREL}
\eeq

\noindent If the $g_V$ term receives contributions only through the combination
$\Gamma_{\mu\nu}$ the relation $f_V = 2g_V$ (many times discussed in the
literature \cite{EQUI,PP,ENJL}) remains valid.  We summarize the two
classes discussed above:

\begin{eqnarray}
{\hbox{Class I}}&:& Z_V, M_V, f_V, g_V~ =~ 2~b\times 2d \nonumber\\
{\hbox{Class II}}&:& H_V, I_V, g_V~ =~ 3~b\times 1d,
\lab{CLASSES}
\end{eqnarray}

\noindent where $g_V$ belongs to both classes by virtue of the identity
(\ref{GVREL}).  Class I receives contribution from a two operators
diagram (n=2), which has the insertion of two boson fields at most,
while the class II receives in general contribution from terms with n=2
and n=3 of the loop expansion.

\subsection{The Leading contributions}

The leading contributions to the parameters in eq. (\ref{CLASSES}) are
obtained by the $\delta_0$ insertion for n=2 and n=3 of the perturbative
expansion. The class I receives contribution from n=2, while the class
II receives contribution from n=3. The diagrams are the following:

\begin{eqnarray}
n=2&\Rightarrow& 1~b\times 1~b    \nonumber\\
n=3&\Rightarrow& 1~b\times 1~b\times 1~b
\end{eqnarray}

The leading divergent contribution to the vector wave function $Z_V$ in
a momentum cutoff regularization scheme is the following:

\beq
Z_V = {N_c\over 16\pi^2}~2\int_0^1~d\alpha \alpha (1-\alpha )
\ln{\Lambda^2\over s(\alpha )}.
\eeq

\noindent The
values of the mass and the coupling constants are obtained after
imposing the correct normalization of the kinetic term of the vector
Lagrangian, which defines the physical vector field as:

\beq
V_\mu = \sqrt {Z_V}~ W^+_\mu
\lab{PHYS}
\eeq

\noindent The mass of the vector meson is given by the mass term of the ENJL
action of eq. (\ref{ENJLBOS}) rescaled by the wave function:

\beq
M_V^2 = {N_c\over 16\pi^2} \biggl ( {\cutoff^2\over 2G_V}\biggr ){1\over Z_V}.
\eeq

The leading divergent contributions to the five parameters of the vector
Lagrangian are given by:

\begin{eqnarray}
Z_V &=& {N_c\over 16\pi^2}~2\int_0^1~d\alpha \alpha (1-\alpha )
\ln{\Lambda^2\over s(\alpha )}
\nonumber\\
M_V^2&=& {N_c\over 16\pi^2} \biggl ( {\cutoff^2\over 2G_V}\biggr
){1\over Z_V}
\nonumber\\
f_V&=& \sqrt{2}\sqrt{Z_V}
          \nonumber\\
g_V&=&{N_c\over 16\pi^2}\sqrt{2} (1-g_A^2){1\over \sqrt{Z_V}}
\int_0^1~d\alpha \alpha (1-\alpha )
\ln{\Lambda^2\over s(\alpha )}
              \nonumber\\ H_V&=&-i{N_c\over 16\pi^2}g_A^2{1\over
\sqrt{Z_V}}
\int_0^1~d\alpha \alpha (1-\alpha )
\ln{\Lambda^2\over s(\alpha )}.
\end{eqnarray}

\noindent The
function $s(\alpha )$ is equal to $M_Q^2 +\alpha (1-\alpha )l^2$ and
depends explicitely upon the external momentum $l^2$.  If we set it to
zero, we produce the low energy limit of the ENJL model derived in
\cite{ENJL}, where the values of the parameters are the following:

\begin{eqnarray}
Z_V &=& {N_c\over 16\pi^2}~{1\over 3}\ln{\Lambda^2\over M_Q^2}
\nonumber\\
M_V^2&=& {N_c\over 16\pi^2} \biggl ( {\cutoff^2\over 2G_V}\biggr ){1\over Z_V}
\nonumber\\
f_V&=& \sqrt{2 Z_V}
          \nonumber\\
g_V&=&{N_c\over 16\pi^2}{\sqrt{2}\over 6} (1-g_A^2){1\over \sqrt{Z_V}}
\ln{\Lambda^2\over M_Q^2}
              \nonumber\\ H_V&=&-i{N_c\over 16\pi^2}{g_A^2\over
6}{1\over \sqrt{Z_V}}
\ln{\Lambda^2\over M_Q^2}.
\lab{PARZERO}
\end{eqnarray}

\noindent They
coincide with the ones calculated in \cite{ENJL} in the proper time
regularization scheme, where one has to use the expression of the
incomplete Gamma function $\Gamma (0,x={M_Q^2\over \cutoff^2})= -\ln x
-\gamma_E + {\cal O}(x)$ for small values of x.

The leading contributions to the parameters of the vector meson
Lagrangian are all logarithmic.  Furthermore the five parameters are not
all independent.  They can be expressed in terms of three of the input
parameters of the ENJL model:

\beq
x={M_Q^2\over \cutoff^2},~~~~G_V,~~~~~g_A.
\eeq

As we will see in the next section this reduction of the number of
independent parameters is no more valid at next-to-leading order.

\subsection{The Next-to-Leading contributions}

As already discussed, the insertion of higher dimension vertices in the
$1/\cutoff$
expansion generates two types of next-to-leading corrections to the parameters
of the vector meson Lagrangian:

\begin{itemize}
\item
leading log power corrections = ${Q^2\over\cutoff^2}\ln {\cutoff^2\over Q^2}$
\item
power corrections = ${Q^2\over\cutoff^2}\cdot 1$
\end{itemize}

\noindent We restrict to the first type of corrections only which
come from a finite set of diagrams constructed with the insertion of one
$1\over\cutoff^2$ vertex.

Using the formulae in Appendix A one can derive the following results.
The insertion of $\delta (x)$ vertices with no derivatives on quark
fields, i.e. contributing as internal quark loop momenta (Case 1. in
Appendix A), contributes to the logarithms of the vector parameters as
follows:

\begin{eqnarray}
1~ &\Rightarrow& \ln {\cutoff^2\over Q^2}=leading   \nonumber\\
{1\over \cutoff^2}&\Rightarrow& {Q^2\over \cutoff^2}\ln {\cutoff^2\over Q^2}
  \nonumber\\
{1\over \cutoff^4}&\Rightarrow& \biggl ( {Q^2\over \cutoff^2}\biggr )^2
\ln {\cutoff^2\over Q^2}
\end{eqnarray}

\noindent The
insertion of $\delta (x)$ vertices with one derivative on quark fields
(Case 2. in Appendix A) contributes as follows:

\begin{eqnarray}
1~ &\Rightarrow& 0 \nonumber\\ {1\over \cutoff^2}&\Rightarrow& {Q^2\over
\cutoff^2}\ln {\cutoff^2\over Q^2} \nonumber\\ {1\over
\cutoff^4}&\Rightarrow& \biggl ( {Q^2\over \cutoff^2}\biggr )^2
\ln {\cutoff^2\over Q^2}
\end{eqnarray}

\noindent The
insertion of $\delta (x)$ vertices with two derivatives on quark fields
(Case 3. in Appendix A) contributes similarly to the previous case:

\begin{eqnarray}
1~ &\Rightarrow& 0 \nonumber\\ {1\over \cutoff^2}&\Rightarrow& {Q^2\over
\cutoff^2}\ln {\cutoff^2\over Q^2} \nonumber\\ {1\over
\cutoff^4}&\Rightarrow& \biggl ( {Q^2\over \cutoff^2}\biggr )^2
\ln {\cutoff^2\over Q^2}  \nonumber\\
{1\over \cutoff^6}&\Rightarrow& \biggl ( {Q^2\over \cutoff^2}\biggr )^3
\ln {\cutoff^2\over Q^2}
\end{eqnarray}

\noindent More
than two derivatives produce terms at least of order $({Q^2/ \cutoff^2})^2
\ln ({\cutoff^2/ Q^2})$ and are beyond our NPLL corrections
which get contribution only from vertices which are at most
$1/\cutoff^2$ and with at most two derivatives on quark fields.

In order to determine how many independent parameters we are left with
after the inclusion of non-renormalizable interactions (NRI) in the
quark-meson Lagrangian, we analyse the corresponding vertices that give
contribution to the five parameters of the Lagrangian ${\cal L}_V$ at
next-to-leading order.

\noindent For n=2 the sets of pairs $(a,b)$ of vertices
$\{ V_1^a\times V_1^b, V_2^a\times V_2^b ...\}$ contributing to each
independent parameter are the following:

\begin{eqnarray}
Z_V &\Leftrightarrow& \biggl\{ -{i\over 2}\gamma_\mu W^{+\mu} \times
{1\over\cutoff^2}
\biggl [
\gamma_\mu d_\lambda W^{+\mu\lambda} + \gamma_\mu  W^{+\mu\lambda} d_\lambda
+ \gamma_\mu (W^{+\mu} d^\lambda - W^{+\lambda} d^\mu ) d_\lambda \biggr
]
\biggr\}
\nonumber\\
f_V&\Leftrightarrow& \biggl\{
\gamma_\mu \Gamma^\mu \times {1\over\cutoff^2}\biggl [
\gamma_\mu d_\lambda W^{+\mu\lambda} + \gamma_\mu  W^{+\mu\lambda} d_\lambda
+ \gamma_\mu (W^{+\mu} d^\lambda - W^{+\lambda} d^\mu ) d_\lambda \biggr ] ,
\nonumber\\
&&{i\over 2}\gamma_\mu W^{+\mu} \times {1\over\cutoff^2}\biggl [
\gamma_\mu d_\lambda \Gamma^{\mu\lambda} + \gamma_\mu
\Gamma^{\mu\lambda} d_\lambda\biggr ] \biggr\}
\nonumber\\
g_V&\Leftrightarrow& \biggl\{ \{\hbox{those of}f_V\} , ~{i\over
2}\gamma_\mu W^{+\mu} \times {1\over\cutoff^2}
\gamma_\mu [\xi_\nu , \xi^\mu d^\nu - \xi^\nu d^\mu ] ,
\nonumber\\
&&{i\over 2}\gamma_\mu \gamma_5 \xi^\mu \times {1\over\cutoff^2}\biggl [
\gamma_\mu\gamma_5 [\xi_\nu , W^{+\mu\nu}] +
\gamma_\mu\gamma_5 [\xi_\nu , W^{+\mu} d^\nu - W^{+\nu} d^\mu ] \nonumber\\
&& +
\gamma_\mu\gamma_5 [W^+_\nu , \xi^\mu d^\nu - \xi^\nu d^\mu ] \biggr ] \biggr\}
\nonumber\\
H_V &\Leftrightarrow& \biggl\{ {i\over 2}\gamma_\mu W^{+\mu} \times
{1\over\cutoff^2}
\biggl [
\gamma_\mu [\xi_\nu , \xi^{\mu\nu}] + \gamma_\mu [\xi_\nu , \xi^\mu d^\nu
- \xi^\nu d^\mu ] \biggr ] ,
\nonumber\\
&&{i\over 2}\gamma_\mu \gamma_5 \xi^\mu \times {1\over\cutoff^2}
\biggl [
\gamma_\mu\gamma_5 [W^+_\nu , \xi^{\mu\nu}] +
\gamma_\mu\gamma_5 [\xi_\nu , W^{+\mu} d^\nu - W^{+\nu} d^\mu ] \nonumber\\
&& +
\gamma_\mu\gamma_5 [W^+_\nu , \xi^\mu d^\nu - \xi^\nu d^\mu ] \biggr ]
\biggr\}
{}.
\lab{LIST1}
\end{eqnarray}

The diagrams with n=3 at $1/\cutoff^2$ order give contributions to
$g_V$ and $H_V$ only:

\begin{eqnarray}
g_V&\Leftrightarrow& \biggl\{ {i\over 2}\gamma_\mu W^{+\mu} \times
{i\over 2}\gamma_\mu \gamma_5 \xi^\mu \times {1\over\cutoff^2}
\gamma_\mu\gamma_5 (\xi^\mu d^\lambda - \xi^\lambda d^\mu ) d_\lambda ,
\nonumber\\
&&\biggl ( {i\over 2}\gamma_\mu \gamma_5 \xi^\mu \biggr )^2 \times
{1\over\cutoff^2} \biggl [
\gamma_\mu  W^{+\mu\lambda} d_\lambda
+ \gamma_\mu (W^{+\mu} d^\lambda - W^{+\lambda} d^\mu ) d_\lambda \biggr
]
\biggr\}
\nonumber\\
H_V &\Leftrightarrow& \biggl\{ {i\over 2}\gamma_\mu W^{+\mu} \times
{i\over 2}\gamma_\mu \gamma_5 \xi^\mu \times {1\over\cutoff^2}\biggl [
\gamma_\mu \gamma_5
\xi^{\mu\lambda} d_\lambda
+ \gamma_\mu\gamma_5 (\xi^\mu d^\lambda - \xi^\lambda d^\mu ) d_\lambda
\biggr ] ,
\nonumber\\
&& \biggl ( {i\over 2}\gamma_\mu \gamma_5 \xi^\mu \biggr )^2\times
{1\over\cutoff^2}
\gamma_\mu (W^{+\mu} d^\lambda - W^{+\lambda} d^\mu ) d_\lambda \biggr\} .
\lab{LIST}
\end{eqnarray}

\noindent Each diagram has one (or two) leading vertex and one NTL vertex.
Each NTL vertex brings a new coefficient $\beta (\cutoff )$.  We
conclude that at NTL order the five parameters of the vector Lagrangian
are all independent.  Each of them has a dependence upon $Q^2$ of the
form:

\beq
f_i =  \biggl ( 1+\beta_i{Q^2\over \cutoff^2}\biggr )\ln{\cutoff^2\over Q^2} .
\eeq

\subsection {The running of $f_V^2$ and $M_V^2$}

For a detailed evaluation of the NTL contributions we concentrate on two
of the five parameters of the vector Lagrangian: the coupling $f_V$
between the vector meson and the external vector current and the mass
$M_V$. In the next section we will use these two parameters in the
calculation of the vector-vector correlation function.  In the ENJL
model (i.e. at the leading order in the logarithmic expansion) the two
parameters are both expressed in terms of the wave function $Z_V$ as
follows:

\beq
f_V = \sqrt{2 Z_V}~~~~~~~~~~~~~~M_V^2 = {N_c\over 16\pi^2} \biggl (
{\cutoff^2\over 2G_V}\biggr ){1\over Z_V},
\eeq

\noindent where
$Z_V$ is the leading logarithmic contribution to the wave-function

\beq
Z_V=Z_V^l = 2{N_c\over 16\pi^2}\int_0^1~d\alpha~\alpha (1-\alpha )\ln
{\cutoff^2\over s(\alpha )}.
\eeq

The product $f_V^2 M_V^2$ is scale invariant:

\beq
f_V^2 M_V^2 = {N_c\over 16\pi^2} {\cutoff^2\over G_V}
\lab{PROD}
\eeq

\noindent By adding the NPLL corrections, the $f_V$ coupling receives
contributions which are absent for the wave function $Z_V$. The latter
defines the renormalized vector mass $M_V$, once the physical vector
field has been defined through eq. (\ref{PHYS}).

A summary of the pairs of vertices entering the calculations of $f_V$
and $Z_V$ (in the same notation of eq. (\ref{LIST1})), including the
leading contributions, is given by:

\begin{eqnarray}
f_V &\Leftrightarrow& \biggl\{ W^+_\mu\times \Gamma_\mu ,
{}~W^+_\mu\times {1\over \cutoff^2}\biggl (
\beta_\Gamma^1 d^\lambda\Gamma_{\mu\lambda} +
\beta_\Gamma^2\Gamma_{\mu\lambda}d^\lambda
\biggr ) ,\nonumber\\
&& \Gamma_\mu\times{1\over \cutoff^2} \biggl (
\beta_V^1d^\lambda W^+_{\mu\lambda} +
\beta_V^2 W^+_{\mu\lambda}d^\lambda +
\beta_V^3 (d^\lambda W^+_\mu- d^\mu W^+_\lambda )d^\lambda
\biggr ) \biggr\}
\nonumber\\
Z_V &\Leftrightarrow& \biggl\{ W^+_\mu\times W^+_\mu ,
{}~W^+_\mu\times{1\over \cutoff^2} \biggl (
\beta_V^1d^\lambda W^+_{\mu\lambda} +
\beta_V^2 W^+_{\mu\lambda}d^\lambda +
\nonumber\\
&&\beta_V^3 (d^\lambda W^+_\mu- d^\mu W^+_\lambda )d^\lambda
\biggr )\biggr\}  ,
\end{eqnarray}

\noindent where a $\beta^i$ coefficient is explicitely written in
front of each $1/\cutoff^2$ vertex.

Using the formula in Appendix A one gets:

\begin{eqnarray}
f_V&=&
\sqrt{2Z_V} + {N_c\over 16\pi^2}{\sqrt{2}\over 3}
{1\over\sqrt{Z_V}}{Q^2\over\cutoff^2}
\biggl [\sum_{i=1}^2~\beta_\Gamma^i
\int_0^1~d\alpha ~P_i(\alpha )\ln{\cutoff^2\over s(\alpha )}
\nonumber\\
&&-{1\over 2}\sum_{i=1}^3~\beta_V^i
\int_0^1~d\alpha ~P_i(\alpha )\ln{\cutoff^2\over s(\alpha )}
\biggr ]
\nonumber\\
M_V^2&=& {N_c\over 16\pi^2}
 \biggl ( {\cutoff^2\over 2G_V}\biggr ){1\over Z_V},
\end{eqnarray}

\noindent where the wave function $Z_V$ is given by:

\begin{eqnarray}
Z_V&=&  {N_c\over 16\pi^2}{1\over 3}\biggl [6
\int_0^1~d\alpha ~\alpha (1-\alpha )\ln{\cutoff^2\over s(\alpha )}
+ \sum_{i=1}^3~\beta_V^i {Q^2\over\cutoff^2}
\int_0^1~d\alpha ~P_i(\alpha )\ln{\cutoff^2\over s(\alpha )}\biggr ]
\nonumber\\
&\equiv& Z_V^l + {N_c\over 16\pi^2}{1\over 3}\sum_{i=1}^3~\beta_V^i
{Q^2\over\cutoff^2}
\int_0^1~d\alpha ~P_i(\alpha )\ln{\cutoff^2\over s(\alpha )}.
\end{eqnarray}

The $\beta^i_{V,\Gamma}$ coefficients must be determined from
experimental data.  The function $s(\alpha )$ is equal to $M_Q^2+\alpha
(1-\alpha )Q^2$.  The $P_i(\alpha )$ are polinomials in the Feynman
parameter $\alpha$.  Their explicit form can be derived by the formula
in Appendix A and reads:

\begin{eqnarray}
P_1 (\alpha )&=& \alpha (1-\alpha ) \nonumber\\ P_2 (\alpha )&=&
\alpha^2 (1-\alpha ) \nonumber\\ P_3 (\alpha )&=& \alpha^3 (1-\alpha
)-3\alpha^2 (1-\alpha )^2
\end{eqnarray}

They correspond to the three possible classes of contributions we got:
1) one vertex with no derivatives on the quark fields 2) one vertex with
one derivative 3) one vertex with two derivatives.  The dependence upon
$Q^2$ of the quantity $\int_0^1~d\alpha ~P_i(\alpha )\ln ({\cutoff^2/
s(\alpha ))}$ for the different $P_i$ is shown in fig. 3.  The
polinomials $P_2$ and $P_3$ lead up to a sign to the same $Q^2$
dependence.  We will use this result for the construction of the
vector-vector correlation function.

\noindent By
including the NPLL corrections, the product (\ref{PROD}) is given by:

\begin{eqnarray}
f_V^2 M_V^2& = &{N_c\over 16\pi^2} {\cutoff^2\over G_V}
\biggl [ 1+ {N_c\over 16\pi^2}{1\over 3}{1\over Z_V}{Q^2\over \cutoff^2}
\biggl ( 2\sum_{i=1}^2~\beta_\Gamma^i
\int_0^1~d\alpha ~P_i(\alpha )\ln{\cutoff^2\over s(\alpha )}
\nonumber\\
&&-\sum_{i=1}^3~\beta_V^i
\int_0^1~d\alpha ~P_i(\alpha )\ln{\cutoff^2\over s(\alpha )}
\biggr )\biggr ] .
\lab{NOSCALE}
\end{eqnarray}

The presence of the new NTL terms with coefficients $\beta_\Gamma^i$ and
$\beta_V^i$ breaks in general the scale invariance of the product in eq.
(\ref{PROD}).

\section{ Phenomenology of the Vector-Vector correlation function.}
\lab{VV}

The finite set of $1/\cutoff^2$ non-renormalizable quark-meson interactions
generates the running of the parameters of the effective meson Lagrangian in
the Quark-Resonance model.

To estimate the values of the coefficients which enter in the running of
$f_V$ and $M_V^2$ we will focus on the particular channel of the vector
resonance sector, by studying the $Q^2$ behaviour of the vector-vector
correlation function where we can compare our predictions with the
experimental results.  We closely follow the derivation of the 2-point
vector function of ref. \cite{2point}.

\noindent We define the 2-point vector function as:

\beq
\Pi_{\mu\nu}^{V(ab)}(q^2)= i\int ~d^4x~ e^{iqx}
<0\vert T(V_\mu^a(x)V_\nu^b(0)\vert 0>,
\eeq

\noindent where $V_\mu^a(x)$ is the flavoured vector quark current defined as:

\beq
V_\mu^a(x)=\bar{q}(x)\gamma_\mu{\lambda^a\over\sqrt{2}}q(x),
\eeq

\noindent with
$\lambda^a$ the Gell-Mann matrices normalised as
$tr(\lambda^a\lambda^b)= 2\delta^{ab}$.  The Lorentz covariance and
$SU(3)$ invariance imply for the $\Pi_{\mu\nu}^V$ the following
structure:

\beq
\Pi_{\mu\nu}^{V(ab)}(q^2)= (q_\mu q_\nu -g_{\mu\nu}q^2)~\Pi_V^1(Q^2)
\delta^{ab}+
q_\mu q_\nu~ \Pi_V^0(Q^2)\delta^{ab},
\eeq

\noindent where $Q^2=-q^2$, with $q^2$ euclidean.
The $SU(3)_L\times SU(3)_R$ ENJL model gives the low energy prediction
for the invariant functions $\Pi^1_V, \Pi^0_V$ in the chiral limit
(${\cal M}\to 0$) and without the inclusion of chiral loops
\cite{2point}:

\begin{eqnarray}
\Pi^1_V(Q^2)&=&-4(2H_1+L_{10}) + {\cal{O}}(Q^2) \nonumber\\
\Pi^0_V(Q^2)&=&0.
\lab{FIRST}
\end{eqnarray}

The parameters $H_1$ and $L_{10}$ are two of the twelve counterterms
that appear in the non anomalous effective Lagrangian of pseudoscalar
mesons at order $p^4$ in the chiral expansion:

\begin{eqnarray}
{\cal L}_4 &=& ......+ L_{10} tr(U^\dagger F_{\mu\nu}^R UF^{\mu\nu}_L)
+H_1 tr(F_{\mu\nu R}^2+F_{\mu\nu L}^2)\nonumber\\ &=& L_{10} {1\over 4}
(f_{\mu\nu}^{+2}- f_{\mu\nu}^{-2}) + H_1 {1\over 2} (f_{\mu\nu}^{+2}+
f_{\mu\nu}^{-2}),
\lab{L4}
\end{eqnarray}

\noindent where
$f_{\mu\nu}^{\pm}$ are related to the external field-strenght tensors
$F_{\mu\nu}^{R,L}$ through the identity:

\beq
f_{\mu\nu}^{\pm}=\xi F_{\mu\nu}^L\xi^\dagger \pm \xi^\dagger F_{\mu\nu}^R\xi
\eeq

\noindent and

\begin{eqnarray}
F_{\mu\nu}^L &=& \partial_\mu l_\nu - \partial_\nu l_\mu - i[l_\mu ,l_\nu ]
\nonumber\\
F_{\mu\nu}^R &=& \partial_\mu r_\nu - \partial_\nu r_\mu - i[r_\mu
,r_\nu ].
\end{eqnarray}

The leading values of $H_1$ and $L_{10}$ at $Q^2=0$ predicted by the QR model
are:

\begin{eqnarray}
H_1&=& -{1\over 12}{N_c\over 16\pi^2} (1+g_A^2)\ln{\cutoff^2\over M_Q^2}
+{\hbox{finite terms}}\nonumber \\
L_{10}&=& -{1\over 6}{N_c\over 16\pi^2} (1-g_A^2)\ln{\cutoff^2\over M_Q^2}
+{\hbox{finite terms}}.
\end{eqnarray}

\noindent The combination $2H_1+L_{10}$ is free from finite contributions.

The Vector-Vector correlation function allows to explore a sector of the
QR model which is free from the effects of the axial-pseudoscalar mixing
(i.e. the parameter $g_A$).  Indeed, the $g_A^2$ dependence is
introduced by the $f_{\mu\nu}^-$ part of the invariant terms, which in
turn depends on the $\xi_\mu$ physical field because of the identity
$f_{\mu\nu}^-=\xi_{\mu\nu}$.  The vector two-point function gets
contribution only from the $f_{\mu\nu}^{+}$ terms and therefore the
parameters $H_1$ and $L_{10}$ will only enter in a combination
independent of $g_A$.  The combination that appears in front of the
$f_{\mu\nu}^{+2}$ term in the Lagrangian (\ref{L4}) is the following:

\beq
{1\over 4}(2H_1+L_{10}) = -{N_c\over 16\pi^2}{1\over 12}
\ln {\cutoff^2\over M_Q^2}
\eeq

\noindent and
contributes as in eq. (\ref{FIRST}) to the two-point vector correlation
function.  As was pointed out in \cite{2point}, the vector resonance
exchange also contributes to the $Q^2$ dependence of the $\Pi^1_V(Q^2)$
function. The diagram with one vector meson exchange in fig. 4
represents the vector resonance contribution to the $\Pi_V^1(Q^2)$. The
total result is:

\beq
\Pi^1_V(Q^2)=-4(2H_1+L_{10}) -2{f_V^2Q^2\over M_V^2+Q^2},
\eeq

\noindent which
includes the contribution at $Q^2=0$ from the genuine one quark-loop
diagram (first term) and the contribution from the vector resonance
exchange (second term). In this approximation the parameters $f_V$ and
$M_V$ are the values at $Q^2=0$ predicted by the ENJL model, i.e. they
are generated by the single quark-loop diagrams with the insertion of
leading vertices in the $1/\cutoff$-expansion (see fig. 5).

 In the ENJL model \cite{ENJL} at $Q^2=0$ the following relation holds:

\beq
(2H_1+L_{10})(Q^2=0) = -{f_V^2\over 2}(Q^2=0)
\lab{ZERO}
\eeq

\noindent so
that the $\Pi_V^1(Q^2)$ function predicted by the ENJL model can be
rewritten in a VMD way

\beq
\Pi^1_V(Q^2)=2{f_V^2M_V^2\over M_V^2+Q^2},
\lab{ENJLZERO}
\eeq

\noindent where the parameters

\beq
f_V^2 = {N_c\over 16\pi^2}{2\over 3}\ln {\cutoff^2\over M_Q^2},~~~~~~~~
M_V^2 = {3\over 2} {\cutoff^2\over G_V} {1\over \ln {\cutoff^2\over
M_Q^2}}
\eeq

\noindent are the values at $Q^2=0$ predicted by the ENJL model.

The authors of \cite{2point} have resummed all quark-bubble diagrams in
fig. 6 with the insertion of the leading 4-quark effective vertex with
coupling $G_V$.  In the VMD representation of eq.  (\ref{ENJLZERO}), the
$Q^2$ dependent contributions coming from the n-loop diagrams can be
reabsorbed in the running of the vector parameters $f_V(Q^2)$ and
$M_V^2(Q^2)$, which are completely determined in terms of the ENJL
parameters. The result quoted in \cite{2point} is the following:

\beq
\Pi^1_V(Q^2)=2{f_V^2(Q^2)M_V^2(Q^2)\over M_V^2(Q^2)+Q^2},
\lab{ENJLRUN}
\eeq

with

\begin{eqnarray}
f_V^2(Q^2) &=& {N_c\over 16\pi^2}~4\int_0^1d\alpha~\alpha (1-\alpha )
\ln {\cutoff^2\over M_Q^2+\alpha (1-\alpha )Q^2}
\nonumber\\
M_V^2(Q^2)&=& {\cutoff^2\over 4G_V}{1\over
\int_0^1d\alpha~\alpha (1-\alpha )
\ln {\cutoff^2\over M_Q^2+\alpha (1-\alpha )Q^2}}.
\lab{ENJLRUN2}
\end{eqnarray}

In the formula (\ref{ENJLRUN2}) we kept only the leading logarithmic
contribution of the expansion of the incomplete Gamma function $\Gamma
(0,x)\simeq -\ln x -\gamma_E +{\cal O}(x)$ appearing in the calculation
of ref. \cite{2point}.

In this case the product $f_V^2(Q^2)M_V^2(Q^2)$ remains scale invariant.

\subsection{$\Pi_V^1(Q^2)$ from the QR model}

The full $Q^2$ dependence of the vector-vector function can be extracted
from the bosonized generating functional. In this case pure fermion
vertices are absent and in particular the 4-fermion vertex with coupling
$G_V$ is replaced by the q-q-V vertex plus a vector mass term, as shown
in fig. 7.

At the one quark-loop level the couplings $H_1, L_{10}, f_V$ and the
mass $M_V$ get NTL logarithmic corrections as we have shown in paragraph
4.3.  The combination $2H_1 +L_{10}$ gets NTL contributions from a
subset of the NTL vertices ($1/\cutoff^2$) which give contribution to
the coupling $f_V$.  The pairs of vertices according to the notation of
eq. (\ref{LIST1}) are:

\begin{eqnarray}
f_V &\Leftrightarrow& \biggl\{ W^+_\mu\times {1\over \cutoff^2}\biggl (
\beta_\Gamma^1 d^\lambda\Gamma_{\mu\lambda} +
\beta_\Gamma^2\Gamma_{\mu\lambda}d^\lambda
\biggr ) , \nonumber\\
&&\Gamma_\mu\times{1\over \cutoff^2} \biggl (
\beta_V^1d^\lambda W^+_{\mu\lambda} +
\beta_V^2 W^+_{\mu\lambda}d^\lambda +
\beta_V^3 (d^\lambda W^+_\mu- d^\mu W^+_\lambda )d^\lambda
\biggr )\biggr\}
\nonumber\\
2H_1+L_{10}&\Leftrightarrow& \biggl\{
\Gamma_\mu\times {1\over \cutoff^2}\biggl (
\beta_\Gamma^1d^\lambda\Gamma_{\mu\lambda} +
\beta_\Gamma^2\Gamma_{\mu\lambda}d^\lambda
\biggr )\biggr\} .
\lab{RUN1}
\end{eqnarray}

Because of the presence of independent unknown coupling constants the
running of the two quantities $f_V^2/2$ and $2H_1+ L_{10}$ is not {\em a
priori} the same. There are two possible solutions at $Q^2\neq 0$:

\begin{itemize}
\item
The running with $Q^2$ of the two parameters can be different, while
their values at $Q^2=0$ are related through the identity (\ref{ZERO}).
In this case the coefficients $\beta_V^i$ and $\beta_\Gamma^i$ of the
NTL logarithmic corrections are not constrained.

\item
The relation (\ref{ZERO}) has to be scale invariant.  This puts a
constraint on the coefficients of the NTL logarithmic corrections to
$f_V^2/2$ and $2H_1+L_{10}$, $\beta_V^i$ and $\beta_\Gamma^i$.

\end{itemize}

The second solution appears to hold in resonance models and under the
saturation hypothesis formulated in \cite{Tens}. For kinematical reasons
the CV model is the only vector model which does not generate the
saturation of the $L_i,H_i$ counterterms of the ${\cal L}_4$ Lagrangian
through vector resonance exchange.  In the ENJL model the saturation is
replaced by the direct contribution of one loop of quarks.  Other vector
models \cite{Tens} saturate the relation (\ref{ZERO}) without the
inclusion of quark-loops contribution. By construction the saturation by
resonance exchange holds at the resonance scale ($Q^2=M_V^2$). If we
require a) the equivalence of the vector models (including the
quark-loops contribution in the case of the CV model) and b) the
validity of the saturation hypothesis, which in fact is experimentally
well verified, we conclude that the relation (\ref{ZERO}) has to be
scale invariant.

\noindent Let us see if this ans\"atz is satisfied by the coefficients
$\beta^i_{V,\Gamma}$.

\noindent The values of the two parameters of eq. (\ref{ZERO}),
including the NPLL corrections, can be deduced by using the formulas in
Appendix A and by inserting the list of vertices of eq. (\ref{RUN1}):

\begin{eqnarray}
{f_V\over \sqrt{2}}&=&\sqrt{Z_V} + {N_c\over 16\pi^2}{1\over 3}
{1\over\sqrt{Z_V}}{Q^2\over\cutoff^2}\biggl [
\sum_{i=1}^2~\beta_\Gamma^i
\int_0^1~d\alpha ~P_i(\alpha )\ln{\cutoff^2\over s(\alpha )}
\nonumber\\
&&-{1\over 2}\sum_{i=1}^3~\beta_V^i
\int_0^1~d\alpha ~P_i(\alpha )\ln{\cutoff^2\over s(\alpha )}
\biggr ]
\nonumber\\
-(2H_1+L_{10})&=& Z_V^l+{N_c\over 16\pi^2}{2\over 3}{Q^2\over\cutoff^2}
\sum_{i=1}^2~\beta_\Gamma^i
\int_0^1~d\alpha ~P_i(\alpha )\ln{\cutoff^2\over s(\alpha )}
\nonumber\\
&=&Z_V +{N_c\over 16\pi^2}{1\over 3}{Q^2\over\cutoff^2}
\biggl [2\sum_{i=1}^2~\beta_\Gamma^i
\int_0^1~d\alpha ~P_i(\alpha )\ln{\cutoff^2\over s(\alpha )}\nonumber\\
&&-\sum_{i=1}^3~\beta_V^i
\int_0^1~d\alpha ~P_i(\alpha )\ln{\cutoff^2\over s(\alpha )}\biggr ] ,
\end{eqnarray}

where the wave function $Z_V$ at one loop is given by:

\begin{eqnarray}
Z_V&=&  {N_c\over 16\pi^2}{1\over 3}\biggl [6
\int_0^1~d\alpha ~\alpha (1-\alpha )\ln{\cutoff^2\over s(\alpha )}
+ \sum_{i=1}^3~\beta_V^i {Q^2\over\cutoff^2}
\int_0^1~d\alpha ~P_i(\alpha )\ln{\cutoff^2\over s(\alpha )}\biggr ]
\nonumber\\
&\equiv& Z_V^l + {N_c\over 16\pi^2}{1\over 3}\sum_{i=1}^3~\beta_V^i
{Q^2\over\cutoff^2}
\int_0^1~d\alpha ~P_i(\alpha )\ln{\cutoff^2\over s(\alpha )}.
\end{eqnarray}

If we compare the running of the two terms of the relation (\ref{ZERO})
up to the NPLL order, we have:

\begin{eqnarray}
{f_V^2\over 2}&=&Z_V + {N_c\over 16\pi^2}{1\over 3}{Q^2\over\cutoff^2}
\biggl [ 2\sum_{i=1}^2~\beta_\Gamma^i
\int_0^1~d\alpha ~P_i(\alpha )\ln{\cutoff^2\over s(\alpha )}\nonumber\\
&&-\sum_{i=1}^3~\beta_V^i
\int_0^1~d\alpha ~P_i(\alpha )\ln{\cutoff^2\over s(\alpha )}
\biggr ]
\nonumber\\
-(2H_1+L_{10})&=& Z_V +{N_c\over 16\pi^2}{1\over 3}{Q^2\over\cutoff^2}
\biggl [2\sum_{i=1}^2~\beta_\Gamma^i
\int_0^1~d\alpha ~P_i(\alpha )\ln{\cutoff^2\over s(\alpha )}\nonumber\\
&&-\sum_{i=1}^3~\beta_V^i
\int_0^1~d\alpha ~P_i(\alpha )\ln{\cutoff^2\over s(\alpha )}\biggr ] .
\lab{RUN2}
\end{eqnarray}

\noindent
They have the same running in $Q^2$ including the NPLL corrections.

$\Pi_V^1(Q^2)$ can be written as follows:

\beq
\Pi_V^1(Q^2) = -4(2H_1+L_{10})(Q^2)-{2f_V^2(Q^2)Q^2\over M_V^2(Q^2)+Q^2} .
\eeq

By using the property that the running of the two parameters in eq.
(\ref{RUN2}) is the same (at least up to the NPLL order) the following
expression holds:

\beq
\Pi_V^1(Q^2) = { 2f_V^2(Q^2)M_V^2(Q^2)\over M_V^2(Q^2)+Q^2},
\lab{FINAL}
\eeq

\noindent where the running of $f_V^2$ and $M_V^2$ is given by:

\begin{eqnarray}
f_V^2&=& 2 Z_V + {N_c\over 16\pi^2}{2\over 3}{Q^2\over\cutoff^2}
\biggl [ 2\sum_{i=1}^2~\beta_\Gamma^i
\int_0^1~d\alpha ~P_i(\alpha )\ln{\cutoff^2\over s(\alpha )}\nonumber\\
&&-\sum_{i=1}^3~\beta_V^i
\int_0^1~d\alpha ~P_i(\alpha )\ln{\cutoff^2\over s(\alpha )}
\biggr ]
\nonumber\\
M_V^2&=&{N_c\over 16\pi^2} \biggl ( {\cutoff^2\over 2G_V}\biggr ){1\over
Z_V}
\nonumber\\
Z_V&=&  {N_c\over 16\pi^2}{1\over 3}\biggl [6
\int_0^1~d\alpha ~\alpha (1-\alpha )\ln{\cutoff^2\over s(\alpha )}
+ \sum_{i=1}^3~\beta_V^i {Q^2\over\cutoff^2}
\int_0^1~d\alpha ~P_i(\alpha )\ln{\cutoff^2\over s(\alpha )}\biggr ] .
\nonumber\\ &&
\lab{FINALR}
\end{eqnarray}

The infinite resummation of quark bubbles considered in ref.
\cite{2point} corresponds to replacing in the vector contribution the
one quark-loop dressed propagator as shown in fig. 8.

The set of diagrams corresponding to the full two-point vector
correlation function predicted by the QR model is shown in fig. 9.

\subsection{ Determination of $\Pi_V^1(Q^2)$ at NTL order}

The real part of the invariant $\Pi$ function is related to its
imaginary part through a standard dispersion relation

\beq
Re\Pi_V^1(Q^2) = \int_0^\infty ds~ { {1\over\pi}Im\Pi_V^1(s)\over s+Q^2}.
\eeq

For a review on QCD spectral Sum rules and the calculation of QCD
two-point Green's functions see \cite{SUMRULE}.

For our analysis we choose the channel of the hadronic current with the
$\rho$ meson quantum numbers ($I=1, J=1$) $J_\mu^\rho =
1/\sqrt{2}(\bar{u}\gamma_\mu u -\bar{d}\gamma_\mu d)$.  The imaginary
part of $\Pi_V^1$ is experimentally known in terms of the total hadronic
ratio of the $e^+e^-$ annihilation in the isovector channel defined as
follows:

\beq
R^{I=1}(s)={\sigma^{I=1} (e^+e^-\to hadrons)\over \sigma (e^+e^-\to
\mu^+\mu^-)}
\eeq

The following dispersion relation holds \cite{shifman,VV}:

\beq
Re \Pi_V^1(Q^2) = {2\over 12\pi^2}\int_0^\infty ds { R^{I=1}(s)\over
s+Q^2}
\eeq

We have performed a comparison between the QR model parametrization of
the vector 2-point function in the isovector channel, valid in the
energy region $0<Q^2<\cutoff^2$, and the prediction obtained from a
modelization of the experimental data on $e^+e^-\to hadrons$ \cite{EXP}.
For a determination of the function $\Pi_V^1(Q^2)$ in the high $Q^2$
region (i.e. beyond the cutoff $\cutoff$) see \cite{HIGH}.

We adopted the following parametrization of the experimental hadronic
isovector ratio:

\beq
R^{I=1}(s) ={9\over 4\alpha^2} {\Gamma_{ee}\Gamma_\rho\over
(\sqrt{s}-m_\rho )^2 +{\Gamma^2\over 4}} + {3\over 2} \biggl (
1+{\alpha_s(s)\over \pi}\biggr )
\theta (s-s_0).
\lab{MODEXP}
\eeq

This is a generalization of the one proposed in ref. \cite{shifman},
where the rho meson width corrections have not been included.
$\Gamma_{ee} = 6.7 \pm 0.4$ KeV is the $\rho\to e^+e^-$ width and
$\Gamma_\rho = 150.9\pm 3.0$ is the total widht of the neutral $\rho$
\cite{PDG}.
We used the leading logarithmic approximation for $\alpha_s(s)$:

\beq
\alpha_s(s) = {12\pi\over 33-2n_f}~{1\over \log (s/ \Lambda_{QCD}^2)}.
\eeq

The modelization (\ref{MODEXP}) includes a dependence of the $\rho$
channel upon the $\rho$ width and the contribution from the continuum
starting at a threshold $s_0=1.5~ GeV^2$ \cite{shifman}.  For the
running of $\alpha_s$ we used a value of 260 MeV for $\Lambda_{QCD}$,
according to the average experimental value $\Lambda_{QCD}^{(4)} = 260
^{+54}_{-46}$ MeV \cite{PDG} and with $n_f=4$ flavours.

The results are practically insensitive to the $\alpha_s$ running
corrections and our leading log approximation turns out to be adequate.

The Vector Green's function in the QR model has been parametrized in
eqs.  (\ref{FINAL}, \ref{FINALR}).  To extract information on the
$\beta_\Gamma^i, \beta_V^i$ coefficients of the NTL logarithmic
corrections we made a best fit of the first derivative of the 2-point
function coming from the modelization (\ref{MODEXP}) of the experimental
data:

\beq
\Pi^\prime (Q^2)_{exp} = -{2\over 12\pi^2}\int_0^\infty ds { R^{I=1}(s)
\over (s+Q^2)^2},
\eeq

\noindent where
the derivative of the VV function in the QR model is given by:

\beq
\Pi^\prime (Q^2)_{QR}={ \biggl [
2f_V^{2\prime} \biggl ( 1+{Q^2\over M_V^2}\biggr ) - 2 {f_V^2\over
M_V^2} \biggl ( 1-Q^2 {M_V^{2\prime}\over M_V^2}\biggr )
\biggr ] \over  \biggl ( 1+{Q^2\over M_V^2}\biggr )^2 }.
\lab{POWERP}
\eeq

\noindent We have used
$M_Q = 265$ MeV for the IR cutoff and $\Lambda_\chi = 1.165$ GeV for the
UV cutoff, determined by a global fit in ref. \cite{ENJL}.

The fit has been done in the region: $0.5<Q<0.9$ GeV. At lower momenta
the NPLL corrections $Q^2/\cutoff^2 \ln (\cutoff^2/Q^2)$ we are
considering compete with the otherwise suppressed logarithmic
corrections $M_Q^2/\cutoff^2 \ln (\cutoff^2/Q^2)$ proportional to
$M_Q^2$.  At higher momenta we are sufficiently near to the cutoff to
require the inclusion of higher order contributions.

The full set of NPLL coefficients consists of the three coefficients
$\beta_V^i$ and the two coefficients $\beta_\Gamma^i$. As we observed in
section 4.3 the polinomials $P_2(\alpha )$ and $P_3(\alpha)$ give rise
to the same $Q^2$ dependence.  We reduced correspondingly the number of
free coefficients in the fit.

We have performed the best fit with four free parameters: $\beta_V^1,\beta_V^2,
\beta_\Gamma^1, \beta_\Gamma^2$.

\noindent In fig. 10 we show the $Q^2$ behaviour of the derivative
of the experimental 2-point function, the data from the best fit, and
the derivative of the ENJL prediction with quark-bubbles resummation of
eqs. (\ref{ENJLRUN}, \ref{ENJLRUN2}) and including only the logarithmic
contributions to the $Q^2$ running of the parameters.  The values of the
coefficients are:

\begin{eqnarray}
{N_c\over 16\pi^2}\beta_\Gamma^1& =& -0.045\pm 0.001~~~~ {N_c\over
16\pi^2}\beta_\Gamma^2 =-0.034\pm 0.002 \nonumber\\ {N_c\over
16\pi^2}\beta_V^1& =& -0.072\pm 0.009~~~~ {N_c\over 16\pi^2}\beta_V^2 =
-0.035\pm 0.01.
\lab{FIT}
\end{eqnarray}

The terms with polinomials $P_1$ and $P_2$ give almost the same
contribution.  The $\chi^2$ of the fit has been defined as $\sum_i
(\Pi^\prime_{i} -
\Pi^{\prime exp}_i)^2/\sigma_i^2$, where we defined the $\sigma_i$ assuming a
$10\%$ of uncertainty on the experimental data: $\sigma_i = 0.1~
\Pi^{\prime exp}_i(Q^2)$.  A $\chi^2/n.d.f. = 5\cdot 10^{-2}$ has been
obtained.  At energies lower than 0.4 GeV the derivative starts to be
sensitive to corrections proportional to the infrared cutoff $M_Q = 265$
MeV, while at energies higher than 0.9 GeV becomes sensitive to higher
order corrections.  The ENJL prediction differs by roughly a $40\%$ from
the experimental curve at 0.8 GeV.

The invariant function $\Pi_V^1(Q^2)$ has been obtained by requiring a
matching with the ENJL function at $Q = M_Q$:

\beq
\Pi_V^1(Q^2) = \Pi_V^{ENJL}(Q^2)~ \theta (M_Q^2-Q^2) +
\int_{M_Q^2}^{Q^2}~ {d\Pi_V^{Fit}\over dQ^{\prime 2}} dQ^{\prime 2}
{}~\theta (Q^2-M_Q^2).
\lab{MATCH1}
\eeq

\noindent The
$\Pi_V^1(Q^2)$ function obtained with the values (\ref{FIT}) and with
the matching of eq. (\ref{MATCH1}) is plotted in fig. 11 and compared
with the ENJL prediction of eq. (\ref{ENJLRUN}) (i.e. including the
resummation of linear chains of quark bubbles and including only
logarithmic corrections).  The difference between the two curves reaches
a $30\%$ at 0.7 GeV.

The inclusion of gluons in the ENJL model makes worse the agreement with
the experimental data.

The modelization of (\ref{MODEXP}) does not include the higher $I=1,J=1$
resonance states with $\rho$ quantum numbers $\rho (1450), \rho (1700)$.
There is no measurement at present of their leptonic width.  The
addition of more resonance states increases the difference between the
two curves. The sensitivity to the continuum threshold value $s_0$ of
$R^{I=1}(s)$ is contained inside a $10\%$ of variation in the range
$s_0=1.5\div 4~GeV^2$. The practical insensitivity to large variations
of the $\Lambda_{QCD}$ parameter, due to the smallness of the
contributions involving $\alpha_s$, has been also verified.

A last point concerns the numerical values used for the IR cutoff $M_Q =
265$ MeV and the UV cutoff $\Lambda_\chi = 1.165$ GeV. They come from a
global fit as explained in detail in ref. \cite{ENJL}. The data used as
inputs are not truly $Q^2=0$ quantities, while the fitted parameters are
the truly $Q^2=0$ values predicted by the ENJL model.  Higher order
corrections in $1/\cutoff^2$, in the scalar sector, can in fact induce
corrections to the mass-gap equation and as a consequence to the
numerical value of the IR cutoff $M_Q$.  A change in the scalar
parameters can modify the prediction for the mass of the scalar
resonance given by the ENJL model. We are currently investigating on
this point.

The analysis of the vecto-vector Green's function shows how a sizable
magnitude of NPLL corrections can be estimated from the data.
Correlations in other channels which are experimentally less accessible
could be estimated by QCD lattice simulations which could be used to fix
the parameters of the effective Lagrangian.
\vspace{1truecm}

\section{ Conclusions.}

Effective quark models inspired to the old Nambu-Jona Lasinio model
\cite{NJL} have proven to be a promising tool to describe low energy
hadronic interactions.  In this type of models the hadron fields are
introduced through the bosonization of the effective quark action. The
effective meson Lagrangian comes from the integration over the quarks
and gluons degrees of freedom.

The simplest model one can construct is the so called ENJL model
\cite{ENJL}, where only the lowest dimension effective quark operators
are included, leading in the $1/\cutoff$ and $1/N_c$ expansions.

As we have shown in detail, the ENJL model correctly predicts the value
of the parameters of the effective meson Lagrangian in the zero energy
limit.  In this limit the model is noticeably more predictive with
respect to the usual effective meson Lagrangian approach
\cite{GL1,GL2,Tens,PP}.  As an example, the twelve counterterms of the
effective pseudoscalar meson Lagrangian at order $p^4$ in the chiral
expansion together with the parameters of the chiral leading effective
resonance Lagrangian are all expressed in terms of only three input
parameters of the NJL model: $G_S, G_V$ and $\cutoff$.  Adding gluon
corrections to order $\alpha_s N_c$ introduces ten more unknown
constants which can be estimated in terms of a single unkown parameter
$g$
\cite{ENJL}.

Nevertheless, the ENJL model is not able to describe the behaviour of
the low energy hadronic observables at $Q^2\neq 0$.

We indicate a systematic way to get predictions on the behaviour of the
hadronic observables in the whole low energy range of $Q^2$ (i.e.
$0<Q^2<\cutoff^2$) which could provide a bridge between the
non-asymptotic and the asymptoptic regime of QCD.

The Quark-Resonance model formulated in this work is based on the
observation that higher dimension n-quark effective interactions give
relevant contributions to the values of the low energy hadronic
observables at $Q^2
\neq 0$.

We have shown that higher dimension operators produce next-to-leading
power - leading log corrections of the type $(Q^2/\cutoff^2) \ln
(\cutoff^2/Q^2)$ to the parameters of the effective meson Lagrangian and
corrections without logarithms of order $(Q^2/\cutoff^2)$. The former
are produced by a finite set of $1/\cutoff^2$ terms, while the latter
arise from an infinite tower of higher dimension operators.

We have focused our attention on the first class of contributions, which
are assumed to be dominant for values of $Q^2$ above the IR cutoff
$M_Q^2$ and below the UV cutoff $\cutoff^2$.

The $Q^2$ behaviour of the low energy observables can be well
reproduced.

\noindent We have shown explicitely how the next-to-leading power -
leading log corrections enter the calculation of the two-point vector
Green's function.  In the $I=1, J=1$ channel we were able to fix the
four coefficients of these corrections through a fit to the experimental
data on the $e^+e^-\to hadrons$ cross section.  The comparison with the
ENJL prediction of ref. \cite{2point} provides evidence for a
quantitative relevance of the next-to-leading terms in the $1/\cutoff$
expansion in the $Q^2$ dependence of the hadronic observables throughout
the intermediate $Q^2$ region.

\vspace{3.truecm}
{\bf Acknowledgements}
\vspace{1.truecm}

E. P. would like to thank Eduardo de Rafael and Johan Bijnens for
interesting and stimulating discussions and for their hospitality at the
Centre de Physique Th\'eorique - Luminy (Marseille) and at the NORDITA
Institute - Copenhagen.

\appendix
\section{Effective Potential calculation: n=2}

The formula to calculate a generic contribution for n=2 in Euclidean
space is the following:

\beq
{1\over 2}\int\int~dx~dy~Tr~\int~{d^4k\over (2\pi )^4} e^{ik(x-y)}
{1\over i\hat{k}+M_Q}\delta (y)
\int~{d^4q\over (2\pi )^4} e^{iq(y-x)}{1\over i\hat{q}+M_Q}\delta (x),
\eeq

\noindent where $Tr$ is the trace over Dirac, colour and flavour indices.

It corresponds to a one quark-loop diagram with two insertions of the
operator $\delta (x)$ as defined in eqs.  (\ref{DELTATOT} and
\ref{DELTA0}).

Defining $l\equiv k-q$ and introducing the Feynman parameter $\alpha$,
the formula reduces to:

\begin{eqnarray}
&&-{1\over 2}\int\int~dx~dy~\int~{d^4l\over (2\pi )^4} e^{il(x-y)}
\int_0^1~d\alpha \int~{d^4k^\prime\over (2\pi )^4}
{1\over [k^{\prime 2} +\alpha (1-\alpha )l^2 +M_Q^2 ]^2 }
\cdot\nonumber\\ && Tr~\biggl \{ [i(\hat{k}^\prime +\alpha \hat{l})-M_Q]
\delta (y) [i(\hat{k}^\prime -(1-\alpha ) \hat{l})-M_Q] \delta (x)
\biggr \} .
\end{eqnarray}

We give here the final formula for the contributions diverging
logarithmically with the cutoff $\cutoff$ obtained with the insertion of
three different forms of the local operator $\delta (x)$.  These are the
only calculations needed to obtain the corrections to the parameters of
the vector meson Lagrangian generated by the insertion of one
next-to-leading vertex $1/\cutoff^2$ and one leading vertex.

\begin{itemize}
\item
Case 1.$~~~~~~~~~$ $\delta (y)=\gamma_\mu (\gamma_5)\delta^\mu (y)~~~~~~
\delta (x)=\gamma_\mu (\gamma_5)\delta^\mu (x)$

\begin{eqnarray}
\Gamma_{log}& =& -{1\over 2}{N_c\pi^2\over (2\pi )^8}\int\int
{}~dx~dy\int~d^4 l~e^{il(x-y)} (l_\mu l_\nu -g_{\mu\nu}l^2)
\nonumber\\
&&tr [\delta^\mu (y)\delta^\nu (x)] 8 \int_0^1~d\alpha~\alpha (1-\alpha
)
\ln{\Lambda^2\over s(\alpha )},
\lab{GLOG}
\end{eqnarray}

\noindent where
$tr$ is the trace over the flavour indices of the $\delta (x)$ matrices
and $s(\alpha )=M_Q^2+\alpha (1-\alpha )l^2$.

\noindent Expression (\ref{GLOG}) can be simplified to:

\begin{eqnarray}
\Gamma_{log}& =& {1\over 2}{N_c\pi^2\over (2\pi )^4}\int\int
{}~dz~dy ~\delta^4(z) \biggl [ \partial_z^\mu \partial_z^\nu - g^{\mu\nu}
\partial_z^2\biggr ]
\nonumber\\
&&tr [\delta^\mu (y)\delta^\nu (z+y)] 8 \int_0^1~d\alpha~\alpha
(1-\alpha )
\ln{\Lambda^2\over s(\alpha )} \nonumber\\
& =& {1\over 2}{N_c\pi^2\over (2\pi )^4}\int ~dy~
tr \biggl [
\delta^\mu (y) ( \partial_\mu \partial_\nu - g_{\mu\nu}\partial^2 )
\delta^\nu (y)\biggr ] \nonumber\\
&&8 \int_0^1~d\alpha~\alpha (1-\alpha )\ln{\Lambda^2\over s(\alpha )}
\end{eqnarray}

\item
Case 2.$~~~~~~~~~$ $\delta (y)=\gamma_\mu (\gamma_5)\delta^\mu (y)~~~~~~
\delta (x)=\gamma_\mu (\gamma_5)\delta^{\mu\lambda} (x)d_\lambda$

\begin{eqnarray}
\Gamma_{log} &=& -{1\over 2}{N_c\pi^2\over (2\pi )^8}\int\int ~dx~dy
\int~d^4 l~e^{il(x-y)} il_\lambda (l_\mu l_\nu -g_{\mu\nu}l^2)
\nonumber\\
&&tr \biggl [ \delta^\mu (y)\delta^{\nu\lambda} (x)\biggr ]
8\int_0^1~d\alpha~ \alpha^2 (1-\alpha )\ln{\Lambda^2\over s(\alpha )}
\nonumber\\
&=&-{1\over 2}{N_c\pi^2\over (2\pi )^4}\int ~dy~
tr \biggl [
\delta^\mu (y) \partial_\lambda
( \partial_\mu \partial_\nu - g_{\mu\nu}\partial^2 )
\delta^{\nu\lambda} (y)\biggr ] \nonumber\\
&&8 \int_0^1~d\alpha~\alpha^2 (1-\alpha )\ln{\Lambda^2\over s(\alpha )}.
\end{eqnarray}

\item
Case 3.$~~~~~~~~~$ $\delta (y)=\gamma_\mu (\gamma_5)\delta^\mu (y)~~~~~~
\delta (x)=\gamma_\mu (\gamma_5)(\delta^\mu (x)d^\lambda-
\delta^\lambda (x)d^\mu )d_\lambda$

\begin{eqnarray}
\Gamma_{log} &=& {1\over 2}{N_c\pi^2\over (2\pi )^8}\int\int ~dx~dy
\int~d^4 l~ e^{il(x-y)}
tr \biggl [\delta^\mu (y)\delta^\nu (x)\biggr ] \nonumber\\
&&~4~\int_0^1~d\alpha~\ln{\Lambda^2\over s(\alpha )}
\biggl \{
\biggl [\alpha^3 (1-\alpha )-3\alpha^2 (1-\alpha )^2\biggr ]
l^2(l_\mu l_\nu -g_{\mu\nu}l^2)  \nonumber\\
&&+{3\over 2} \biggl [ {3\over 2}\alpha^2 (1-\alpha )^2
-\alpha^3 (1-\alpha )\biggr ] g_{\mu\nu}l^4
\biggr\} .
\lab{CASE3}
\end{eqnarray}

\end{itemize}

The last term proportional to $g_{\mu\nu}l^4$ in eq. (\ref{CASE3}) does
not contribute to the logarithmically divergent part of the integral and
one obtains:

\begin{eqnarray}
\Gamma_{log}&=&{1\over 2}{N_c\pi^2\over (2\pi )^4}\int ~dy~
tr \biggl [
\delta^\mu (y) \partial^2
( \partial_\mu \partial_\nu - g_{\mu\nu}\partial^2 )
\delta^\nu (y)\biggr ] \nonumber\\
&&4 \int_0^1~d\alpha~
\biggl [\alpha^3 (1-\alpha )-3\alpha^2 (1-\alpha )^2\biggr ]
\ln{\Lambda^2\over s(\alpha )}.
\end{eqnarray}

We have not included logarithmic terms proportional to the IR cutoff
mass $M_Q$.

\vfill\eject
\centerline {{\bf TABLE CAPTIONS}}
\vskip2.truecm

\begin{description}
\item[1)] Parity and Charge Conjugation transformation properties of the
quark bilinears.

\item[2)] Parity and Charge Conjugation transformation properties of the
fundamental fields of the effective meson theory.

\end{description}
\vfill\eject

\centerline {{\bf FIGURE CAPTIONS}}
\vskip2.truecm

\begin{description}

\item[1)]
The QCD diagram with one gluon exchange generates an effective 4-quark
interaction vertex.

\item[2)]
A quark-loop diagram with at least one meson field as external leg. The
integration over quarks ( and gluons) produces the vertices of the
effective meson Lagrangian.  Double lines are resonances, dotted lines
are pions and wavy lines are the external currents.

\item[3)]
The integrals $\int_0^1~d\alpha P_i (\alpha ) \ln (\cutoff^2/s(\alpha
))$ which occur in the NTL logarithmic corrections to the effective
meson Lagrangian are shown as a function of $\sqrt{Q^2}$. The three
polinomials correspond to the three cases of Appendix A.

\item[4)]
The vector meson exchange at the tree level gives the $Q^2$ dependent
term of the vector-vector correlation function in a vector resonance
model. The ENJL model gives the prediction (\ref{PARZERO}) for the
parameters of the vector meson Lagrangian.

\item[5)]
The running of $f_V$ with $Q^2$ generated by the QR model: the full
circle indicates the insertion of a leading (${\cal{O}}(1)$) or a
next-to-leading (${\cal{O}}(1/\cutoff^2$) vertex in the one quark-loop
diagram.

\item[6)]
The resummation of n-quark bubble diagrams which gives the full $Q^2$
dependence of the vector-vector correlation function in the ENJL model
of ref. \cite{2point}. They contain the insertion of the leading 4-quark
vector vertex with coupling $G_V$.

\item[7)]
The 4-quark vector vertex of the fermion action with coupling $G_V$ is
replaced by the sum of the q-q-vector vertex and the mass term of the
vector field in the bosonized action.

\item[8)]
The "dressed" vector meson propagator is given by the resummation of n
quark-loop diagrams which are leading in the $1/N_c$ expansion.

\item[9)]
The full vector two-point function as predicted by the QR model which we
remind is developed at the leading order in the $1/N_c$ expansion.  The
vector meson propagator of the second term is defined in fig. 8.

\item[10)] The derivative of the experimental vector-vector function
$-d\Pi_V^1(Q^2)/dQ^2$ (solid line), the fitted curve of the QR model
(dashed line) and the prediction of the ENJL model including
quark-bubble resummation and the logarithmic contributions in the
incomplete Gamma functions $\Gamma (0, x)$ \cite{2point} (dot-dashed
line) are shown as a function of $\sqrt{Q^2}$. The fit has been
performed in the region $0.5\geq \sqrt{Q^2}\leq 0.9$ GeV.

\item[11)] The invariant function $\Pi_V^1(Q^2)$
(dashed line) is obtained from the fitted derivative of fig. 10 by
imposing the matching with the ENJL function at $Q=M_Q$. The ENJL
prediction of eq. (\ref{ENJLRUN} (full line) is also shown.  Gluon
contributions have not been included.

\end{description}
\vfill\eject

\begin{table}
\centering
\begin{tabular}{|c|c c|}\hline
  & & \\ & P & C \\ \hline & & \\ $V_\mu$ & $\eps (\mu )$ & $-V_\mu^T$
\\ & & \\ $A_\mu$ & $-\eps (\mu )$ & $A_\mu^T$ \\ & & \\ $\sigma$ &
$\sigma$ & $\sigma^T$ \\ & & \\ $\Gamma_\mu$ & $\eps (\mu )$ &
$-\Gamma_\mu^T$ \\ & & \\ $\xi_\mu$ & $-\eps (\mu )$ & $\xi_\mu^T$ \\ &
& \\ $f_{\mu\nu}^{\pm}$ & $\pm\eps (\mu )\eps (\nu )$ &
$\mp{f_{\mu\nu}^\pm}^T$\\ & & \\ $\chi_\pm$ & $\pm\chi_\pm$ &
$\chi_\pm^T$ \\ & & \\ \hline
\end{tabular}
\end{table}
\vskip0.7truecm
\centerline{Table 1.}
\vfill\eject

\begin{table}
\centering
\begin{tabular}{|c|c c|}\hline
  & & \\ & P & C \\ \hline & & \\ $\bar{Q}Q$ & $+$ & $+$ \\ & & \\
$\bar{Q}\gamma_5Q$ & $-$ & $+$ \\ & & \\ $\bar{Q}\gamma_\mu\gamma_5Q$ &
$-\epsilon (\mu )$ & $(\bar{Q}\gamma_\mu\gamma_5Q)^T$ \\ & & \\
$\bar{Q}\gamma_\mu Q$ & $\epsilon (\mu )$ & $-(\bar{Q}\gamma_\mu Q)^T$
\\ & & \\ $\bar{Q}\sigma_{\mu\nu}Q$ & $\epsilon (\mu )\epsilon (\nu )$ &
$-(\bar{Q}\sigma_{\mu\nu} Q)^T$ \\ & & \\
$\bar{Q}\sigma_{\mu\nu}\gamma_5 Q$ & $\sim \epsilon^{\mu\nu\alpha\beta}
V_{\alpha\beta}$ & \\ & & \\ \hline
\end{tabular}
\end{table}
\vskip0.7truecm
\centerline{Table 2.}
\vfill\eject


\begin{thebibliography}{99}
\bibitem{GL1}
J. Gasser and H. Leutwyler, {\em Ann. Phys.} {\bf 158} (1984) 142.

\bibitem{GL2}
J. Gasser and H. Leutwyler, {\em Nucl. Phys.} {\bf B250} (1985) 465,
517.

\bibitem{Fuji}
T. Fujiwara et al., {\em Progr. Theor. Phys.}{\bf 73} (1985) 926.

\bibitem{Bando}
M. Bando, T. Kugo and K. Yamawaki, {\em Phys. Rep.} {\bf 164} (1988)
 217.

\bibitem{EFFL}
J. Bijnens, A. Bramon and F. Cornet, {\em Z. Phys.}{\bf C 46} (1990)
 599.

\bibitem{Tens}
G. Ecker et al., {\em Nucl. Phys.} {\bf B321} (1989) 311.

\bibitem{PP}
E. Pallante and R. Petronzio, {\em Nucl. Phys.} {\bf B396} (1993) 205.

\bibitem{ENJL}
J. Bijnens, C. Bruno and E. de Rafael, {\em Nucl. Phys.} {\bf B390}
(1993) 501.

\bibitem{NJL}
Y. Nambu and G. Jona-Lasinio, {\em Phys. Rev.} {\bf 122} (1961) 345;

Y. Nambu and G. Jona-Lasinio, {\em Phys. Rev.} {\bf 124} (1961) 246;

D. J. Gross and A. Neveu, {\em Phys. Rev.} {\bf D10} (1974) 3235.

\bibitem{NC}
G. 't Hooft, {\em Nucl. Phys.} {\bf B72} (1974) 461;

E. Witten, {\em Nucl. Phys.} {\bf B160} (1979) 57;

\bibitem{Andrianov}
A. A. Andrianov and V. A. Andrianov,
{\em Int. J. of Mod. Phys.} {\bf A8} (1993) 1981.

\bibitem{2point}
J. Bijnens, E. de Rafael and H. Zheng, Preprint CERN-TH 6924/93,
CTP-93/P2917, NORDITA-93/43 N,P.

\bibitem{bijnens}
J. Bijnens and E. de Rafael, {\em Phys. Lett.} {\bf B273} (1991) 483.

\bibitem{MANYNJL}
T. Eguchi, {\em Phys. Rev.} {\bf D14} (1976) 2755;

D. Ebert and H. Reinhard, {\em Nucl. Phys.} {\bf B271} (1986) 188;

V. Bernard, R. L. Jaffe and U.-G. Meissner,
{\em Nucl. Phys.} {\bf B308} (1988) 753;

M. Schaden et al., {\em Nucl. Phys.} {\bf B339} (1990) 595;

M. Wakamatsu, {\em Annals of Phys.} {\bf 193} (1989) 287.

\bibitem{boh}
D. Espriu, E. de Rafael and J. Taron, {\em Nucl. Phys.} {\bf B345}
(1990) 22;

V. A. Novikov, M. A. Shifman, A. I. Vainshtein and V. I. Zakharov,
{\em Fortschr. Phys.} {\bf 32} (1984) 585.

\bibitem{Peris}
S. Peris and E. de Rafael, {\em Phys. Lett.} {\bf B309} (1993) 389.

\bibitem{Georgi}
A. Manohar and H. Georgi, {\em Nucl. Phys.} {\bf B234} (1984) 189;

\bibitem{PRADES}
J. Prades, preprint CPT-93/P.2871.

\bibitem{EQUI}
G. Ecker et al., {\em Phys. Lett.} {\bf B223} (1989) 425.

\bibitem{shifman}
M. A. Shifman, A. I. Vainshtein and V. I. Zakharov, {\em Nucl. Phys.}
{\bf B147} (1979) 385, 447.

\bibitem{VV}
S. Narison, "Exotic Mesons of QCD", preprint PM/87-51, October 1987,
Montpellier, Talk presented the VI LEAR Workshop 6-13 September 1987,
Villars-sur-Ollon (Switzerland).

\bibitem{SUMRULE}
S. Narison, "QCD Spectral Sum Rules", World Scientific Lecture Notes in
Physics, Vol. 26;

E. G. Floratos, S. Narison and E. de Rafael, {\em Nucl. Phys.} {\bf
B155} (1979) 115;

C. Becchi, S. Narison, E. de Rafael and F. J. Yndur\`ain, {\em Z. Phys.}
{\bf C8} (1981) 335;

S. Weinberg, {\em Phys. Rev. Lett.} {\bf 18} (1967) 507.


\bibitem{HIGH}
A. De R\'ujula and H. Georgi, {\em Phys. Rev.} {\bf D13} (1976) 1296;

G. J. Gounaris and J. J. Sakurai, {\em Phys. Rev. Lett.} {\bf 21} (1968)
244;

F. J. Yndur\'ain, {\em Phys. Lett.} {\bf 63B} (1976) 211;

R. Shankar, {\em Phys. Rev.} {\bf D15} (1977) 755;

T. Appelquist and H. Georgi, {\em Phys. Rev.} {\bf D8} (1973) 4000;

A. Zee, {\em Phys. Rev.} {\bf D8} (1973) 4038;

D. J. Broadhurst, {\em Phys. Lett.} {\bf 101B} (1981) 423;

H. D. Politzer, {\em Phys. Rev. Lett.} {\bf 30} (1973) 1346.

\bibitem{EXP}

J.-E. Augustin et al.,  {\em Phys. Rev. Lett.} {\bf 34} (1975) 764;

D. Benaksas et al., {\em Phys. Lett.} {\bf 39B} (1972) 289;

S. I. Dolinsky et al., {\em Phys. Rep.} {\bf 202} (1991) 99.

\bibitem{PDG}
Review of Particle Data, {\em Phys. Rev.} {\bf D45} (1992) 1.

\end{thebibliography}
\end{document}